\documentclass[aps,prb,twocolumn,showpacs,floatfix,superscriptaddress]{revtex4}

\usepackage{graphicx}
\usepackage{amssymb}
\usepackage{amsfonts}
\usepackage{amsmath}
\usepackage{dcolumn}

\usepackage{soul}

\usepackage{makecell}

\usepackage{color}
\usepackage{stackrel}
\usepackage{latexsym}

\usepackage[normalem]{ulem}
\usepackage{times}
\usepackage{bm}
\usepackage{hyperref}
\usepackage{accents}

\def\bk{{\bf k}}
\def\sku{{Pt$_4$Ge$_{12}$}\;}

\definecolor{capri}{rgb}{0.0, 0.75, 1.0}

\begin{document}

\title{First Principles Study of Electronic Structure and Fermi Surface\\
 in Rare-earth Filled Skutterudites $R$Pt$_4$Ge$_{12}$ }

\author{Gheorghe Lucian Pascut}
\affiliation{Department of Physics \& Astronomy, Rutgers University, Piscataway, NJ O8854, USA}
\affiliation{MANSiD Research Center and Faculty of Forestry, Stefan Cel Mare University (USV), Suceava 720229, Romania}

\author{Michael Widom}
\affiliation{Department of Physics, Carnegie Mellon University, Pittsburgh, PA  15213, USA}

\author{Kristjan Haule}
\affiliation{Department of Physics \& Astronomy, Rutgers University, Piscataway, NJ O8854, USA}

\author{Khandker F. Quader}
\affiliation{Department of Physics, Kent State University, Kent , OH 44242, USA}

\date{\today}

\begin{abstract}
Experiments on rare-earth filled skutterudites demonstrate an intriguing array of thermodynamic, transport and superconducting properties, and bring to fore theoretical challenges posed by f-electron systems. First principle calculations based density functional theory and its extensions for strongly correlated systems such as the Hubbard U correction, provide valuable information about electronic structure that can be used to understand experiments. We present a comprehensive study of the electronic structure and Fermi surface of a series of rare earth filled skutterudites, R\sku (where R = La, Ce, Pr), aimed at shedding light on: consequences of progressive increase of f-orbital occupancy in the series; the effects of the Hubbard parameter U; the Fermi surfaces, band structures and densities of states. The calculated Fermi surfaces may be relevant to the question of multi-band versus single-band superconductivity. Computed densities of states qualitatively explain the available resonant photoemission spectroscopy experiments, and (together with available specific heat measurements) provide estimates of the effective masses. We also show the existence of pseudogaps in the total density of states which may be relevant for the thermoelectric properties of these systems.
\end{abstract} 
  
%\pacs{}

\maketitle

\section{Introduction}
\label{sec_Introduction}

\begin{figure*}[!htb]
\includegraphics[width=0.85\linewidth]{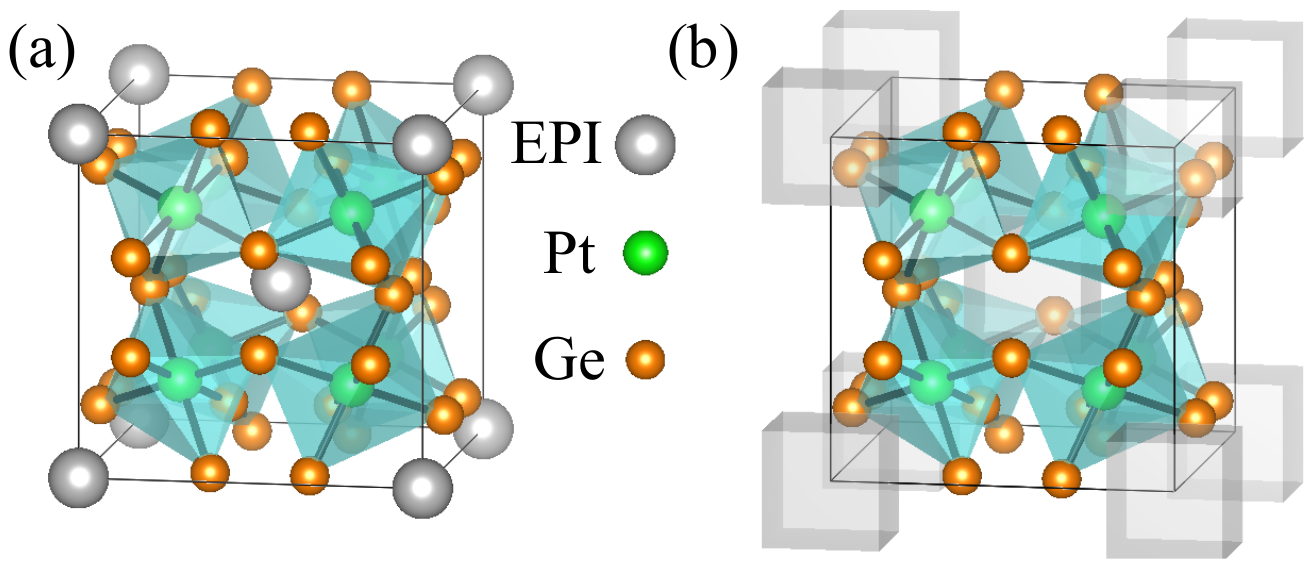}
\caption{(Color online) \textbf{Crystal Structure:} (a) filled skutterudite and (b) skutterudite. EPI stands for electropositive ion. The voids, mentioned in the text, are shown by the cubic transparent shapes.}
\label{fig:struct}
\end{figure*}

Filled skutterudites are a class of materials that exhibit complex crystal structures ~\cite{CS_Jeitschko,R_Gumeniuk_CS_2010}, and correspondingly complex electronic ground states, such as conventional BCS-type~\cite{BCS_KL_Maple} and unconventional superconductivity~\cite{GLP5_Unconventional_Superconductivity,MaisuradzePRL09,UncSuper_PhysRevB.82.024524}, non-Fermi liquid behavior~\cite{non_fermi_liquid1,non_fermi_liquid2}, anomalous metal-to-insulator transition~\cite{MIT_PhysRevLett.79.3218}, multipolar ordering~\cite{multipolar_ordering}, topological insulator state ~\cite{TI1_PhysRevB.85.165125,TI2_PhysRevLett.106.056401}, Kondo lattice behavior~\cite{BCS_KL_Maple}, valence fluctuations ~\cite{TodaJPS08}, heavy fermion behavior~\cite{HFermionB_Maple_PhysRevB.65.100506,HFermionB_Maple,HFermionB_Maple_MAPLE2005830}, various magnetically ordered states~\cite{MAGStates_MAPLE2007182,MAGStates_SATO2009749,MaplePRB16,MAGStates_PhysRevB.73.144409}, etc. They are also promising materials for next generation thermoelectric applications.~\cite{TherMat1_Morelli_1995,TherMat2_Sales_1997_PhysRevB.56.15081,TherMat3_Nolas_1998_PhysRevB.58.164,TherMat4_Sales1325_1996}

In this paper, using first principles density functional theory (DFT) and its extensions for strongly correlated systems, we address open questions about the Fermi surface  topology and its s, p, d, f character in the recently synthesized R\sku filled skutterudites with R = La, Ce or Pr. Our study is driven by the large body of experimental work on the  R\sku, that demonstrate an array of novel properties~\cite{TherMat1_Morelli_1995,non_fermi_liquid1,MIT_PhysRevLett.79.3218,MAGStates_MAPLE2007182,UncSuper_PhysRevB.82.024524,BCS_KL_Maple}. Existing experimental results (summarized in Section~\ref{sec_Experimental_Background}) point to the need for comparative study of electronic properties in R\sku compounds. Such a study would be useful to understand similarities and differences between these materials and the consequence of these for macroscopic properties. Despite some existing density of states (DOS) calculations~\cite{GumeniukPRL08,GumeniukJP11,ChandraPhilmag16}, a comprehensive study across the range of these compounds including spin-orbit coupling (SOC) and the Hubbard parameter $U$ is still lacking. As pointed out in Ref.~\onlinecite{Zhang-GumenPRB13}, such calculations may shed light on possible multi-band nature of superconductivity and the origin of multiple and/or complicated pairing mechanisms. Thus our paper reports a systematic and comparative study of the R\sku electronic structure. 

The complex crystal structure of the filled skutterudites (shown in Fig.~\ref{fig:struct}a) is stabilized by adding an electropositive ion to the empty voids of the skutterudites (see Fig.~\ref{fig:struct}b). Besides stabilizing the crystal structure it turns out that the electropositive ion (which in our work is La, Ce or Pr) also gives rise to novel electronic states~\cite{R_Gumeniuk_CS_2010} and improved thermoelectric properties~\cite{TherMat4_Sales1325_1996}. Considering the complex crystal structure and the fact that usually $f$-electron systems pose challenges for DFT methods~\cite{DFT_problems_PhysRevB.89.205109,DFT_problems_Petit_2016} we confirm the robustness of our main results, by carrying out an extensive study under various theoretical scenarios such as DFT+SOC and DFT+SOC+U with the $f$-electrons treated as core or valence electrons. Among our goals is to understand the changes in the electronic structure and Fermi surface (FS) topology and its character, as we progress from La to Ce and Pr, possessing, respectively, 0, 1, and 2 $f$-electrons in their pure elemental states.

We find that independently of the theoretical approximation used in our calculations: (I) multiple large Fermi surfaces exists in R\sku, a feature similar to that in MgB$_2$ where multi-band superconductivity is found~\cite{MgB2_Xi_2008,MgB2_PhysRevLett.86.4656}; (II) The large FS have anistropic orbital character on the FS, a feature similar to the FS in Pb where multi-band superconductivity is also found~\cite{Pd_PhysRevB.75.054508,Pd_PhysRev.186.397}; (III) in all R\sku compounds the states at $E_F$ are dominated by Ge-$p$ states; (IV) $f$ states are always present at $E_F$ for Ce\sku and Pr\sku independently of the values of the Hubbard parameter U; (V) SOC lifts band degeneracy, shifts bands and affects the topology of the Fermi surfaces; (VI) DOS show deep pseudogaps above $E_F$ which may have implications for thermoelectric properties. In addition, our DFT+SOC+U calculations qualitatively explain features observed in photoemission experiments, while revealing a need for improved treatment of the $f$-electrons systems. We also find that the deduced values of the mass enhancement are not insignificant signaling the existence of modest correlations in these materials.

The paper is organized as follows. In Section~\ref{sec_Experimental_Background}, we give a brief overview of the experimental behavior of filled skutterudites.  Section~\ref{Sec_Methods} is devoted to a discussion of the methods used for our calculations and the different theoretical scenarios considered. This is followed by Section~\ref{sec_Results} where we present the results for the rare-earth-filled skutterudites, R\sku  (R=La, Ce, Pr) under different scenarios, namely PBE, PBE + SOC, PBE+SOC +U. Results for band structure, energy bands and FS without the the inclusion of Hubbard like correlation $U$ are presented in Sections ~\ref{sec_Results_A}, ~\ref{sec_Projected_DOS_B}, and ~\ref{sec_3D_Fermi_Surfaces_C} respectively, and the effects of including $U$ on DOS and FS are presented in Section ~\ref{sec_Effects_of_correlations_D}.  In Section~\ref{sec_Effective_mass_E}, based on measurements of low-temperature specific heat, we provide simple estimates of the effective mass enhancements over the band mass, $m^*/m_b$  for the various theoretical scenarios considered. We discuss possible thermoelectric application of the R\sku  in Section~\ref{sec_Thermoelectric_Applications_F}. We end the main text with discussions in Section~\ref{sec_Discussion} and acknowledgments in Section~\ref{sec_Acknowledgements}. We note that we provide additional details in appendices in Section~\ref{sec_Appendix}.

\section{Experimental Background}
\label{sec_Experimental_Background}

Experiments on Ba\sku, Sr\sku,  and R\sku skutterudites reveal a diverse array of interesting properties: while Ba-, Sr-, La-, Pr-\sku  exhibit superconductivity at 5.1, 5.35, 8.3, 7.9 K respectively~\cite{{BauerPRL07,RosnerPRB09},GumeniukPRL08}, Ce\sku  does not show superconductivity, and instead it is believed to demonstrate mixed valence or Kondo lattice behavior, or possibly a behavior in between.~\cite{TodaJPS08,GumeniukJP11} Nd\sku and Eu\sku  are characterized by anitferromagnetic  transitions at 0.67 and 1.7 K respectively~\cite{GumeniukPRL08}. Sm\sku exhibits\cite{GumeniukNewjPhys10} heavy fermion behavior with a large value of the electronic specific heat coefficient ($\gamma \sim$ 450/mol/K$^2$), indicating the possibility of strong correlations in this material.

In the Ba, Sr and La compounds, experiments point to conventional nodeless s-wave BCS pairing; however, sub-linear behavior of the  field-dependent electronic part of specific heat $C_V(H)$ and thermal conductivity $\kappa (H)$ may indicate two equal-sized s-wave superconducting gaps in La-\sku.~\cite{SteglichPRB16,ChandraPhilmag12,ChandraPhilmag16} The nature of pairing in PrPt$_4$Ge$_{12}$ is less settled. Low-T specific heat and muon spin resonance measurements claim evidence for unconventional pairing with point nodes in the gap~\cite{MaisuradzePRL09}. But, several other experiments point to multi-band superconductivity with two superconducting gaps. Lower and upper critical field measurements and photoemission spectroscopy (PES)~\cite{NakamuraJPSJ10,NakamuraPRB12}, and upturn in temperature dependence of upper critical field~\cite{ChandraPhilmag12,ChandraPhilmag16,GurevichPhysica07} indicate multi-band paring. Strong indication of  two-gap superconductivity (as in MgB$_2$~\cite{MgB2_TSUDA200436,MgB2_TSUDA2007126,MgB2_Xi_2008,MgB2_PhysRevLett.86.4656} and Pb~\cite{Pd_PhysRevB.75.054508,Pd_PhysRev.186.397,Pd_PhysRevB.4.1523}) has been found in the analysis of superfluid density in {\it single} crystal Pr\sku~\cite{Zhang-GumenPRB13}, the only single crystal  measurement in this material that we are aware of. Multi-band pairing with one nodeless  and one nodal gap has been inferred~\cite{SinghPRB16} from fitting low-T specific heat data in Pr\sku and Pr$_{1-x}$Ce$_{x}$\sku, while other work based on transport and thermodynamic measurements on Pr$_{1-x}$Ce$_{x}$\sku~\cite{MaplePRB14}, and PrPt$_4$Ge$_{12-x}$Sb$_x$~\cite{MaplePRB16} leave open the possibility of nodal versus multi-band superconductivity. $^{73}$Ge nuclear quadrupole resonance~\cite{KanatakeJPS10} shows a coherence peak below pairing transition temperature, typical of BCS pairing. 

A detailed study of the electronic structure of R\sku may be able to shed some light on the extent to which the Pr 4$f$ electrons play a role in these couplings, and thereby indirectly playing a role in multi-band pairing and gaps. The strong experimental indication of multi-band pairing in Pr\sku, and possibly in La\sku, is suggestive of a complex band structure with several bands crossing the Fermi surface. Therefore a detailed study of the electronic properties in R\sku is necessary.

The R\sku materials studied here do not exhibit long range magnetic order. La\sku is diamagnetic at all temperatures, while the $f$-electrons in Ce\sku and Pr\sku  obey Curie-Weiss behavior at sufficiently high temperatures. For Pr\sku, the magnetic susceptibility starts to exhibit a Curie-Weiss behavior for T $>$100 K~\cite{TodaJPS08} with experimental average effective magnetic moment $\mu_{CW}$  of 3.59$\mu_B$/f.u.~\cite{TodaJPS08} or 3.69$\mu_B$/f.u.~\cite{MaplePRB14,GumeniukPRL08} ($\mu_B$ being Bohr magneton) that are close to that of the free Pr$^{3+}$ ion (3.58 $\mu_B$) with the electronic configuration 4f$^2$ and the $^3$H$_4$ Hund rule ground state multiplet, indicating the presence of local moments on the Pr ions. However, the magnetic susceptibility saturates at low temperatures, indicating a non-magnetic ground state, scenario that is consistent with the crystal field splitting of the degenerate $^3$H$_4$ Hund rule ground state multiplet~\cite{TodaJPS08,GumeniukPRL08}.

In the case of Ce\sku, the magnetic susceptibility has a also Curie-Weiss behavior down to 200 K~\cite{TodaJPS08} with experimental average effective magnetic moment $\mu_{CW}$ of  2.58$\mu_B$/f.u. or 2.51$\mu_B$/f.u.~\cite{GumeniukJP11} that are close to that of the free Ce$^{3+}$ ion (2.54 $\mu_B$) with the electronic configuration 4f$^1$ and the $^2$F$_{5/2}$ Hund rule ground state multiplet, indicating the presence of local moments on the Ce ions. Below 200 K, the susceptibility deviates from Curie-Weiss and exhibits a broad maximum at 80 K,  followed by an upturn below $\sim$20K~\cite{TodaJPS08,GumeniukJP11}. The broad maximum can be interpreted as a characteristic features of intermediate valence~\cite{TodaJPS08} or as indicative of local moment screening with a large characteristic energy~\cite{GumeniukJP11}.

From all R\sku compounds studied here, Ce\sku is the only one where a long range order, of antiferromagnetic type, can be stabilized at low temperatures upon Sb substitution of Ge~\cite{Ce_Mag_PhysRevLett.109.236405}. Due to the absence of local moments in La\sku or the non-magnetic ground state of the Pr ions in Pr\sku, no long range magnetic order is found in La\sku or Pr\sku compounds. 

\begin{figure*}[!htb]
\includegraphics[width=0.85\linewidth]{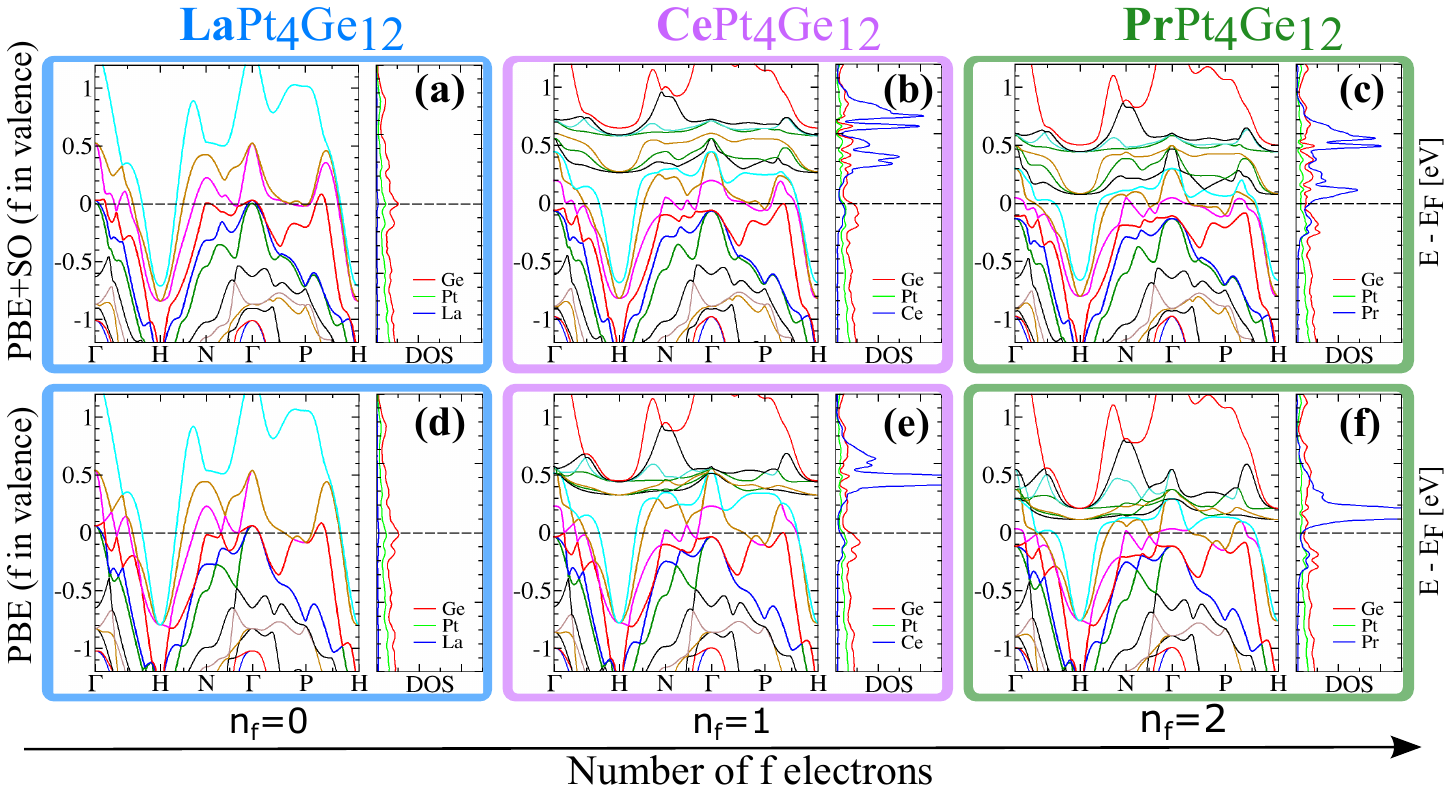}
\caption{(Color online) \textbf{Electronic Band Structure and Atom-Projected DOS} results for R\sku compounds obtained using the VASP code. Each column corresponds to a particular compound whose chemical formula is printed at the top of the column. Each row corresponds to results from calculations using a particular approximation whose label is printed at the beginning of the row.  \textit{On the left side of each panel} we show the electronic band structure (energy in eV on the vertical axis versus momentum on the horizontal axis). Fermi energy is marked by the horizontal dashed line. \textit{On the right side of each panel} we show atom-projected DOS (each unit on the DOS axis represents 1 state/eV/f.u.).} 
\label{Figure2_BANDS}
\end{figure*}

In addition to superconducting, magnetic, and transport properties, filled skutterudites appear promising with respect to their thermoelectric capabilities ~\cite{thermoelectric_capabilities}. Since the capabilities of a thermoelectric material is a consequence of the interplay between the transport quantities such as the Seebeck coefficient S, the electrical resistivity $\rho$, and the thermal conductivity $\kappa$, optimizing these quantities in a material could give rise to a large thermoelectric figure of merit, ZT$\ge$1, that is required for reasonable performance of a  thermoelectric material. Recently it was shown that ZT could be further improved in filled skutterudites by tuning the thermal conductivity based on the so-called Phonon Glass and Electron Crystal (PGEC) concept which was developed for cage-forming structures in general ~\cite{PGEC1,PGEC2,PGEC3}. In addition, related to the materials studied in this work, it has been shown that by substituting Ge by Sb in LaPt$_4$Ge$_{12}$ compound, the Seebeck coefficient and resistivity could be enhanced by an order of magnitude at room temperature ~\cite{thermoelectric_capabilities}, thus increasing the playground to optimize the figure of merit ZT.
 
\begin{figure*}[!htb]
\includegraphics[width=0.85\linewidth]{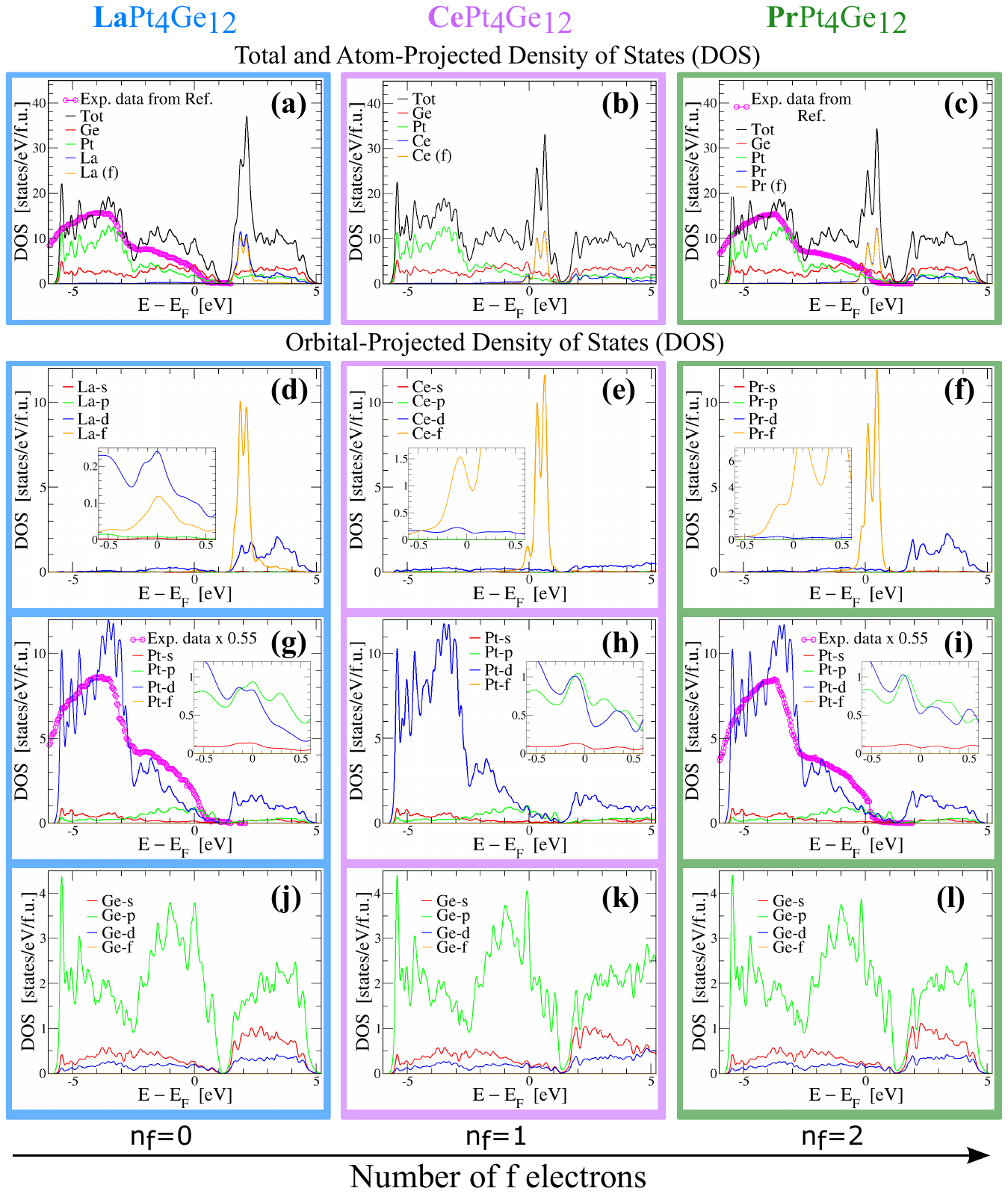}
\caption{(Color online) \textbf{Total, Atom- and Orbital-Projected DOS} results for Re\sku compounds. The results presented in this figure were obtained using the VASP code within the PBE+SO approximation, treating the $f$-electrons as valence electrons.  Each column corresponds to a particular compound. Each panel on the first row corresponds to Total, atom- and $f$-projected DOS. Panels, on the other three rows show orbital-projected DOS for the three distinct atoms making up these compounds. Fermi energy is marked by the zero on the horizontal axis. Some panels enlarge the DOS near the Fermi energy in an inset. The experimental soft X-ray photoemission spectra for LaPt$_4$Ge$_{12}$ and PrPt$_4$Ge$_{12}$  presented in Ref.~\onlinecite{NakamuraJPS10} were digitized and included in our figures for comparison.} 
\label{Figure3_GDOS}
\end{figure*}

\section{Methods}
\label{Sec_Methods}

All our calculations are based on the electronic density functional theory method~\cite{Hohn-Kohn,Kohn-Sham} with the Perdew-Burke-Ernzerhof (PBE) generalized gradient approximation~\cite{PBE} to the exchange-correlation potential.  Calculations are performed on the experimental crystal structures having the Im-3 cubic space group ~\cite{R_Gumeniuk_CS_2010}. 

For each compound, we studied in detail the electronic band structure, DOS and FS under various theoretical scenarios such as PBE, PBE+SOC and PBE+SOC+U. For each of these scenarios the rare-earth $f$-electrons were treated either as valence electrons or core electrons. We include, SOC and U in our calculations due to the fact that they are important energy scales for rare earth elements.

Owing to the experimentally reported absence of magnetism in the R\sku compounds studied in this work, we performed all our calculations in non-spin polarized states by constraining the total spin moment per unit cell to zero. At the same time, we made sure that all our solutions also have the orbital moments equal to zero for each theoretical scenario considered in this work. Since we set the initial atomic moments to zero, every band should obey time reversal symmetry, $E_\sigma({\bf k})=E_{-\sigma}(-{\bf k})$. In addition, our structure has inversion symmetry, so we expect $E_\sigma({\bf k})=E_\sigma(-{\bf k})$ which we checked to be the case here. As a consequence, each FS presented in this paper is doubly degenerate. 

In this work we combine results from VASP and WIEN2k codes, to take advantage of their strengths. We note that WIEN2k is an all electron code (core + valence electrons), while VASP works only with the valence electrons (while the potential of the atom nuclei and the core electrons is replaced by an effective potential known as a pseudopotential). Even though the basis sets in the two codes differ, as long as the basis set in each code is almost complete basis, the results obtained by the two codes should be very similar. We compared  the electronic band structure of the R\sku obtained by the two codes for each theoretical scenario used in this work and in each case we found very similar electronic structure, as expected.

Below, we give specific parameters used for the calculations in each code.

{\bf VASP}: Electronic band structure and DOS calculations utilize the plane-wave based DFT code VASP~\cite{Kresse96} with the all-electron projector augmented wave (PAW) method~\cite{Kresse99}. We take the standard VASP PAW potentials, keeping the $f$-orbitals as valence states (except where we specify $f$ in core). Increasing plane wave energy cutoffs had no discernible impact on the band structure so we apply the VASP defaults. However, we use the full FFT grids consistent with the energy cutoff to represent charge densities so as to avoid wrap-around errors. Electronic $k$-point meshes are increased to $31\times 31\times 31$ in order to converge the electronic DOS, which we evaluate using tetrahedron integration followed by 0.01eV Gaussian smearing.

{\bf WIEN2k}: FS and the associated electronic band structure were performed using the full-potential linearized augmented plane wave (FP-LAPW) method as implemented in the WIEN2k code~\cite{wien2k}. The calculations were performed on a 31 $\times$ 31 $\times$ 31 $k$-point mesh.  The muffin-tin radii were chosen as 2.50, 2.45 and 2.22 Bohr units for the rare earth ions, Pt and Ge, respectively. In order to have accurate calculations we used RmtKmax = 8. The energy which separates the core and the valence states was chosen to be $-$10 Ry. Self-consistent criteria for the energy and charge were ~10$^{- 4}$ Ry and ~10$^{- 4}$ electrons, respectively. All other input parameters were used with their default values.

\section{Results}
\label{sec_Results}

\subsection{Overview of Band Structure}
\label{sec_Results_A}

Fig.~\ref{Figure2_BANDS} presents electronic band structure and atom-projected DOS for PBE and PBE+SOC approximations, where $f$-electrons are treated as valence electrons. For each band structure plot, the $k$-point path that connects special $\bk$ points in the first Brillouin zone (BZ) is defined as in Ref.~\onlinecite{SETYAWAN2010299}: $\Gamma$  $(0,0,0)$, $H$ $(1/2,-1/2,1/2)$, $N$ $(0,0,1/2)$, and $P$ $(1/4,1/4,1/4)$.

Several things may be noted in the set of plots in Fig.~\ref{Figure2_BANDS}: 

(I) From the electronic band structure plots for the R\sku compounds for a given approximation (rows in Fig.~\ref{Figure2_BANDS}), we see that by going from La to Pr (which is equivalent with increasing the number of $f$-electrons, n$_f$, from 0 to 2) new bands appear above the Fermi level (E$_F$) in the energy range (0.2 - 0.6) eV. The new bands are related to the Ce and Pr $f$-electrons, and although they hybridize with the other bands in the system, we see that the bands in the vicinity of the E$_F$ don't change their shape drastically in Ce\sku and Pr\sku  when compared to La\sku, but they are shifted down in energy with respect to the E$_F$. Since the intersections of the bands with the E$_F$ defines points on the 3D FS, by shifting the bands around the E$_F$ will change the topology of the FS. Thus, we expect that the topology and the number of the FS in Ce\sku and Pr\sku to be different from the FS in La\sku.

(II) The atom-projected DOS gives information about the contribution/weight of each atom to the electronic band structure at a given energy. Thus our calculations show that in the La compound, the states at the E$_F$ come mostly from the Pt and Ge atomic states and there are few or no La states. By contrast, in the Ce and Pr compounds, besides the Pt and Ge atomic states, we also have states from the rare-earth elements. As we will discuss later in paper this can have implications for the single- versus multi-band superconductivity in R\sku compounds.

(III) Comparing plots with and without SOC for a given compound, see Fig.~\ref{Figure2_BANDS}, panels (a)-(c) versus (d)-(f), we see that SOC lifts the band degeneracy at special points and along various k-paths in the first BZ and shifts some of the bands by up to a few hundred meV. For example, these effects are especially evident at the $\Gamma$ and H points and along the $\Gamma$-P path. Since SOC has an impact on the electronic band structure, it also has a direct effect on the topology and degeneracy of the FS.

It is well-known that for rare earth elements, the occupied $f$-states can behave as inert localized states in some compounds (and in those cases the $f$-electrons must be treated as core in the theoretical approximations) or the $f$-states can hybridization with the other states in the system (and in those cases the  $f$-electrons must be treated as valence electrons). In Appendix A, we show the electronic band structure and atom projected DOS, for the case where the $f$-electrons are treated as core, thereby eliminating hybridization with the other states in the system. From the calculations with $f$-electrons in core, we learn that for a given approximation, these compounds have almost identical electronic band structure and thus almost the same FS as expected (see rows in Fig.~\ref{Figure2_BANDS_Appendix}). Thus, the theoretical approximations where the $f$-electrons are treated as core electrons can not explain the differences between the La\sku and Ce\sku or Pr\sku compounds. The effects of SOC on the band structure are similar, independently of the way we treat the $f$-electrons (in valence or in core).

\subsection{Projected DOS}
\label{sec_Projected_DOS_B}

In Fig.~\ref{Figure3_GDOS}(a)-(c), we show the total DOS per formula unit of R\sku compounds together with the atom-projected DOS. In addition, in Fig.~\ref{Figure3_GDOS}(d)-(l), we also show the orbital-projected DOS for each atom within the R\sku compounds. These calculations were done using the PBE+SOC approximation within VASP with the $f$-electrons treated as valence electrons. In this figure we also show the comparison of our calculations with the available experimental soft X-ray photoemission spectra~\cite{NakamuraJPS10}. We digitized the experimental data and we plot it on top of our calculated DOS in Fig.~\ref{Figure3_GDOS}(a) and (c). Since the experimental data is in arbitrary units, we scaled it such that we get the best agreement between the experiment and theory in the energy range around -4 eV. We see that when comparing the experimental data with the total DOS per formula unit, by construction we get a good agreement for the energy range around -4 eV, but the agreement is poor for energy closer to the E$_F$. The reason for this discrepancy is the fact that the photoemission spectra were measured at an incident energy of 1.2 KeV, energy where mostly Pt d states were probed, the other states practically being invisible to this probe due to the very small scattering cross-section at this energy~\cite{NakamuraJPS10}. These discrepancies are expected since the total DOS per formula unit is a sum of all states with equal probability without considering the experimental scattering cross-sections. But if we consider the scattering cross-sections, and we compare the photoemission spectra only with the Pt d DOS as shown in Fig.~\ref{Figure3_GDOS}(g) and (i), we see that the agreement between experiment and theory is much better over the full energy range. 

Looking at all orbital projected DOS, see Fig.~\ref{Figure3_GDOS}(d)-(l), we see that the orbital contributions of the s, p and d states at the E$_F$ is similar for all three compounds. The difference is that in the case of Ce and Pr compounds, $f$-states are also present at the E$_F$. This suggests that the electrons forming the Cooper pairs in Pr\sku might have some additional $f$ character besides the $s$, $p$ and $d$, while in La\sku their character is mostly $s$, $p$ and $d$. Based on our results, and making an analogy with the case of Pb ~\cite{Pd_PhysRevB.75.054508}, where the anisotropic contributions of the s, p and d orbital character to the wavefunctions forming the states at the E$_F$ leads to anisotropic electron-phonon coupling and multi-band superconductivity, we suggest the possibility of multi-band superconductivity for Pr\sku and single- or multi-band superconductivity for La\sku. The pseudogaps found in the total DOS of the R\sku compounds, might be relevant to possible improvement of their {\it thermoelectric effects}, as we will discuss later on in the paper.

\subsection{3D Fermi Surfaces}
\label{sec_3D_Fermi_Surfaces_C}

\begin{figure*}[!htb]
\includegraphics[width=0.90\linewidth]{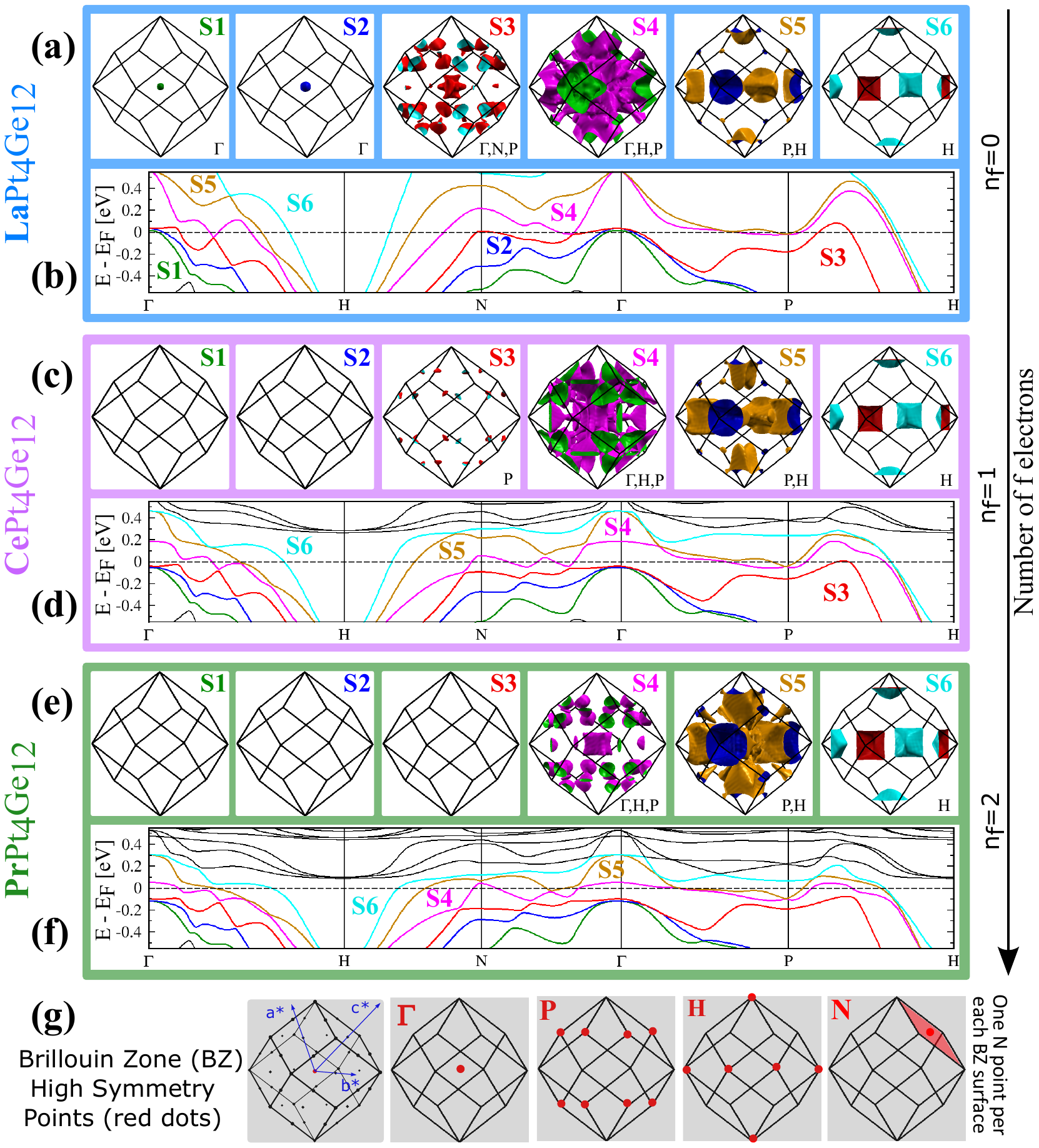}
\caption{(Color online) \textbf{Fermi Surfaces (FS)} for RPt$_4$Ge$_{12}$ compounds obtained by using the Wien2k code with the PBE+SOC approximation and treating the $f$-electrons as valence electrons. Each diagram corresponding to one of the three compounds contains multiple panels showing the FS within the first Brillouin zone (panels a, c and e) and the electronic band structure around the E$_F$ (panels b, d and f). FS and the corresponding bands have the same colors as the symbols Sn (n=1 to 6) that represents them. Panel g shows the reciprocal lattice vectors, a$^*$, b$^*$, c$^*$ and the position within first Brillouin Zone (BZ) of the  special reciprocal points $\Gamma$, P, H, N. On each FS panel, we also print the label of the special reciprocal points where the FS exists.} 
\label{Figure4_FS}
\end{figure*}

In Fig.~\ref{Figure4_FS} we show the 3D FS (labeled by symbols Sn, n=1 to 6) for the R\sku. These FS are calculated in WIEN2k, using the PBE+SOC approximation and treating the $f$-electrons as valence electrons. In addition to the FS, we also show the electronic band structure for each compound in a narrow energy range around the Fermi energy.

As can be seen in Fig.~\ref{Figure4_FS}(a), La\sku has six FS. The hole or electron pocket character can be ascribed to these FS by considering the corresponding band structure in Fig.~\ref{Figure4_FS}(b). S1, S2 and S3 are hole pockets, and S5 and S6 are electron pockets. The character of S4 is not as clear from the band structure, see Fig.~\ref{Figure4_FS}(b). Thus it is interesting to examine the changes of the FS with increasing number of $f$- electrons. With 1 additional electron (compare Ce 4$f^1$ with La 4$f^0$), the Fermi energy increases, and the S1 and S2 hole pockets get filled while the S3 pocket is nearly filled, see Fig.~4(c). With the addition of another electron (the case of Pr 4$f^2$) the S3 Fermi hole pocket is completely filled,  see Fig.~4(e). The upshot is that Ce\sku has four FS and Pr\sku has only three, see Fig.~\ref{Figure4_FS}(c) and (e). Within the PBE+SOC approximation, the three compounds have in common three large FS, namely S4, S5 and S6. As the number of $f$-electrons increases, S4 shrinks in size and hence has dominant hole character, while S5 grows and S6 is hardly affected. The S1-S3 hole pockets present in La\sku and absent in Ce\sku and Pr\sku within the PBE+SOC approximation, show up  within the PBE+SOC+U approximation for both Ce\sku and Pr\sku compounds. 

The orbital character of the FS may influence the nature of superconductivity. It is claimed that Pb  has two Fermi surfaces ~\cite{Pd_PhysRevB.75.054508,Pd_PhysRev.186.397}, one spherical and one more complex. Owing to the anisotropic orbital character of the wave functions ($s$, $p$ and $d$ across the two FS), the electron-phonon coupling varies across the FS and between the two FS, thereby giving rise to multi-band superconductivity. It may also be recalled that in MgB$_2$, multi-band superconductivity is given by two FS, both with $p$ character (one FS is $p_z$ in character and the other is $p_{x,y}$ in character)~\cite{MgB2_TSUDA200436,MgB2_TSUDA2007126,MgB2_Xi_2008,MgB2_PhysRevLett.86.4656,GurevichPhysica07}. If the mechanism for superconductivity in the La and Pr compound is also phonon-mediated, since there are multiple sizable FS as in MgB$_2$, this implies the possibility of multi-band superconductivity in R\sku, as pointed out by experiments. In addition, since the Pr compound is shown to have  $f$ character on the FS (as shown by DOS in Fig.~\ref{Figure3_GDOS}), this could lead to different electron-phonon couplings across the FS, similar to the anisotropic electron-phonon coupling found in Pb. Thus, the electron-phonon couplings may be different in Pr\sku compared to that in the La\sku case, thereby resulting in differences in the type of pairing. In order to ascertain the anisotropic orbital character across the FS, besides looking at the orbital projected DOS which gives an average of the orbital contribution over the all k-points in the BZ, we also looked at the "fat-band" representation~\cite{FatBands} around the E$_F$ which gives a detailed description of the orbital contribution at specific k-points in the BZ. The "fat-band" plots are shown in Appendix B. From the "fat-band" representation, we find anisotropic orbital character to the 3D FS for all the orbital states coming from R, Pt and Ge atoms. 

\subsection{Effects of correlations on DOS and FS}
\label{sec_Effects_of_correlations_D}

\begin{figure}
\includegraphics[width=0.85\linewidth]{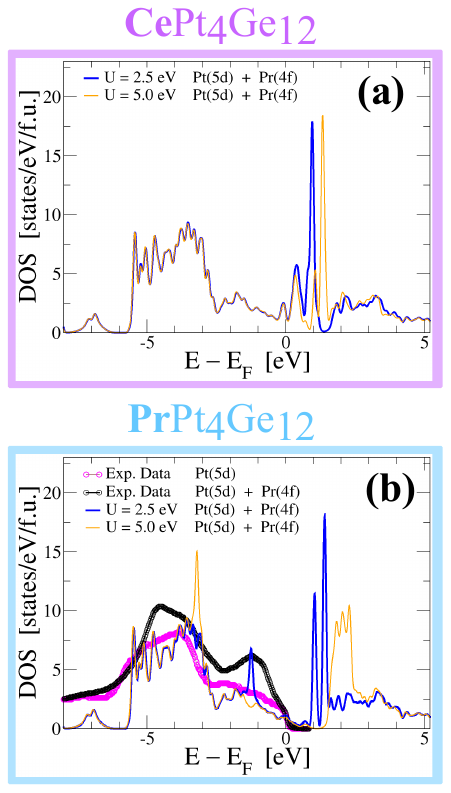}
\caption{(Color online) \textbf{Correlation effects on DOS}. Panels (a) and (b) present results of the PBE+SOC+U approximation for Ce and Pr compounds. The experimental soft X-ray photoemission spectra for PrPt$_4$Ge$_{12}$  presented in Ref.~\onlinecite{NakamuraJPS10} were digitized and included in our figure for comparison. In addition, panel (b) shows the experimental spectra which has contributions mostly from Pt 5d states (magenta symbols) and the experimental spectra which has contributions from both Pt 5d and Pr 4f states (black symbols). For details see Fig. 2 in Ref.~\onlinecite{NakamuraJPS10}.} 
\label{dos-vs-U}
\end{figure}

Photoemission measurements on Pr\sku where the incident energy is scanned through the $f$-resonance of the Pr atoms, reveal the presence $f$ states at E$_F$ and below. For example, see Fig.~\ref{dos-vs-U}b where we show the digitized data from Ref.~\onlinecite{NakamuraJPS10} for two incident energies, 1.2 and 0.9 KeV. For incident energy of 1.2 KeV we probe mostly the Pt $d$ states (magenta curve) and for incident energy of 0.9 KeV, in addition to Pt $d$ states, we also probe the Pr $f$ states (black curve). Thus, the difference between the magenta and black curves in Fig.~\ref{dos-vs-U}b represents the contribution of the Pr $f$ states to DOS.  Besides small contributions at the E$_F$, the Pr $f$ states appear to manifest as two peaks centered around -1 and -4.5 eV which disagrees with our DOS calculations for the PBE+SO approximation shown in Fig.~\ref{Figure3_GDOS}f. In order to explain the experimentally observed $f$ contribution to DOS, shown in Fig.~\ref{dos-vs-U}b we need to go beyond the DFT approximations. Repulsive interactions among the tightly bound $f$-orbitals of the Lanthanide atoms can create electronic correlations that are not properly modeled by our PBE density functional. A first step in correcting this effect is to supplement the DFT Hamiltonian with onsite Coulomb energies~\cite{U_PhysRevB.57.1505,U_PhysRevB.48.16929}, which we take into account using a single effective Hubbard parameter $U$. We take $U=5$ eV, a typical value for Lanthanides, to illustrate the qualitative effect. For comparison, we also explore other values of $U$. 

DOS computed within the PBE+SOC+U with $U$=0 eV shown in Fig.~\ref{Figure3_GDOS} reveal that the $f$-band is partly occupied, with the net $f$ occupation of $n_f$=0.85 for Ce and $n_f$=2.07 for Pr. For $U>0$, occupied and empty $f$ orbitals split proportionately to $U$ for Pr, but the splitting is far weaker for Ce, consistent with the respective values of $n_f$. For example, in Fig.~\ref{dos-vs-U} we show the Pt(d)+Pr(f) DOS for Ce\sku and Pr\sku compounds computed using $U =$ 2.5 and 5.0 eV. As expected, the effect of U is to increase the splitting between the occupied and unoccupied $f$ states, thus moving the theoretical $f$ peak of occupied states in the Pt(d)+Pr(f) DOS, to lower energies. Comparing our theoretical DOS with the experimental data, we find that the position of the theoretical $f$ DOS could explain the experimental peak shape at -1eV for U = 2.5 eV or the peak at -4.5 eV for U = 5.0 eV, but our theory can not explain both experimental peaks at the same time. This suggests that PBE+SOC+U method captures part of the electronic correlation, but in order to explain the experimental findings a theory beyond DFT+$U$ is required. One such theory, where correlation effects are treated more rigorously is the density functional theory + embedded dynamical mean field theory (DFT+eDMFT)~\cite{Haule_prb10,JPSJ_Haule,webpage}.

Since DOS at the E$_F$ are affected by U, the FS of Ce\sku and Pr\sku will also be affected. Since the positions of bands with respect to $E_F$ directly relate to the topology of the FS, we also expect changes in the topology of the FS. While additional details are given in Appendix B (see Fig.~\ref{Ce-bands-U} for Ce\sku and Fig.~\ref{Pr-bands-U} for Pr\sku), we note a few salient points here. The band corresponding to the S6 FS does not change too much with increasing $U$ indicating that it is mostly $s$, $p$ and $d$ in character.  The bands corresponding to S4 and S5 change with increasing $U$, S4 being more affected than S5, suggesting that these FS having stronger $f$ character than S6. To confirm the contribution of $f$ states to the S4 and S5 FS for finite $U$ values, we have also computed the FS for the case where the $f$-electrons are treated as core electrons (in this case the Fermi surface has no $f$ contribution). The fact that the topology of the FS for finite $U$ differs from the case where the $f$-electrons are treated as core electrons (labeled $U~\sim~\infty~eV$ in Appendix B Fig.~\ref{Ce-bands-U} and Fig.~\ref{Pr-bands-U}), confirms once again the presence of $f$ character in the Fermi surface of Ce\sku and Pr\sku, independently of the strength of correlations. In addition, we also learn that the presence or absence of the S1, S2 and S3 FS in Ce\sku and Pr\sku depends on the strength of the correlations. 

\subsection{Effective mass}
\label{sec_Effective_mass_E}

While our calculations show that the rare-earth filled skutterudites have complex band structures and DOS, simple estimates of effective mass can nevertheless be obtained by considering the linear coefficient of the specific heat obtained experimentally in these materials. 
The electronic specific heat 
$C_V (T)$ in metals, for $T<< T_F$ ($T_F$ being the Fermi temperature), is given by:~\cite{Kittel8}
\begin{equation}
C_V = \gamma T = \frac{1}{3} \pi^2 k_B^2 N(E_F) T.
\end{equation}
For a spherical FS with parabolic bands, the linear coefficient of specific heat, $\gamma$, is related to the electronic DOS at the E$_F$ as $N(E_F) = m^*k_F/\hbar^2\pi^2$, with $m^*$ being the single isotropic effective mass $m^*$, and $k_F$ the Fermi momentum. 
Assuming that the theoretical density of states at the E$_F$, $D(E_F)$, can likewise be related to a single band mass $m_b$, we obtain
estimates of the mass enhancement over the band mass, $m^*/m_b$, by comparing the measured $N(E_F)$ (obtained from $\gamma$) and the computed density of states at E$_F$, $D(E_F)$. While in reality, the density of states and effective masses are expected to be anisotropic, thermodynamic measurements, such as specific heat, are averages over FS; hence simple analyses as this may provide a reasonable estimate of $m^*/m_b$.

Table~\ref{tab:summary} summarizes results drawn from the various theoretical scenarios we have considered in this work. In particular we show the calculated total DOS at the E$_F$, D$(E_F)$ and mass enhancement over the band mass, $m^*/m_b$. For a given theoretical scenario, different values of estimated $m^*/m_b$ for each skutterudite correspond to $\gamma$'s obtained from specific heat measurements by different groups. The estimated mass enhancements, $m^*/m_b$, in the range of 2-4, are giving an indication of the correlation strength.

\begin{table*}
\center
\begin{tabular}{|l|ccc|ccc|ccc|cc}
\hline
DFT  scenarios & \multicolumn{3}{c|}{D($E_F$)[states/eV/f.u.] - calc} & \multicolumn{3}{c|}{$m^*/m_b$ - estimate}\\
\hline
& La  & Ce & Pr  & La$^{a,b}$ & Ce$^c$ & Pr$^{d,e,f}$\\
\hline
PBE (f in valence)  & 11.9 & 10.9 & 14.7 & 2.7/2.0 & 4.1 & 2.5/2.1/2.0\\
PBE + SOC & 10.0 &12.2 &16.4 & 3.2/2.4 & 3.7 & 2.3/1.9/1.8\\
PBE + SOC + U ( = 5) & -- & 10.0 & 11.3 & -- & 4.5 & 3.3/2.8/2.6\\
PBE (f in core) & -- & 9.6 & 9.6 & -- & 4.7 & 3.9/3.3/3.1\\
\hline
\end{tabular}
\caption{\label{tab:summary} Theoretical DOS at $E_F$ and the corresponding estimated mass enhancement over the band mass, $m^*/m_b$ (see text for details). For a given compound, different $m^*/m_b$ correspond to different measurements of $\gamma$, the linear coefficient of specific heat $C_V$, in units of mJ/mol K$^2$: $a$: Ref. [\onlinecite{GumeniukPRL08}] - $\gamma$ = 75.8; $b$: Ref. [\onlinecite{SteglichPRB16}] - $\gamma$ = 56; $c$: Ref. [\onlinecite{GumeniukJP11}] - $\gamma$ = 105; $d$: Ref. [\onlinecite{GumeniukPRL08}] - $\gamma$ = 87;  $e$: Ref. [\onlinecite{SinghPRB16}] - $\gamma$ = 73.7; $f$: Ref. [\onlinecite{Zhang-GumenPRB13}] - $\gamma$ = 69 (single crystal).}
\end{table*}

\subsection{Thermoelectric Applications}
\label{sec_Thermoelectric_Applications_F}

There has been sustained interest in the skutterudites as possible next generation
thermoelectric materials. Viability for such application is measured in terms of the figure of merit, $ZT$ (preferably greater than 1) given by
\begin{equation}
  \label{eq:ZT}
ZT = \frac{S^2 \sigma}{\kappa} T
\end{equation}
where $\kappa$ contains electronic and lattice contributions to thermal conductivity,
$\sigma$ is the electrical conductivity, and $S$ the Seebeck coefficient. To achieve $ZT > 1$ requires an optimal interplay of various quantities. 
Our theoretical calculations suggest that these materials are metallic, with a sizable  DOS at $E_F$ and thus a small Seebeck coefficient is expected at room temperature. One of the ways to increase $ZT$, is to increase the Seebeck coefficient which, in simple terms, is given by:~\cite{Ziman72,Ketterson16}
\begin{equation}
  \label{eq:S}
S(T) = \frac{\pi^2 k_B^2 T}{3e} \Big[ \frac{N(E)}{n(E)} + \frac{d \ln \mu(E)}{dE}\Big]_{E=E_F}    
\end{equation}
where $e$, $n(E)$, $N(E)$, $\mu$, and $E_F$ are the charge, carrier density, density of states, mobility, and the Fermi energy, respectively. Thus S depends on the density of states and inversely on the carrier density, as well as the logarithmic derivative of the carrier mobility $\mu$ with energy. In the above derivation of $S(T)$, conductivity $\sigma$, and mobility $\mu$ have been taken to be: $\sigma(E) = e n(E) \mu(E)$ and $\mu = |e| \tau(E)/m^*$, where $\tau$ is the carrier lifetime which can be energy-dependent, and $m^*$ the carrier effective mass. Thus in effect, the second term in $S(T)$ can depend on the energy variation of the lifetime. 
The derivative of $\mu$ accounts for the way in which electron current is distributed in energy. Thus, if $\mu(E)$ is an increasing function 
of energy, a relatively higher proportion of current will be carried by the more energetic carriers, and these will transport a larger amount of heat.  
The above simple result provides a sense of the expected behavior of thermopower for metals at low temperatures $T/T_F << 1$. In a more realistic scenario, one would need to express conductivity in terms of Fermi surface area, and treat density of states and lifetime
in more details.

As was discussed in Ref.~\onlinecite{thermoelectric_capabilities}, large S can be achieved by driving the system close to a metal to insulator transition where the charge carrier density, $n$, diminishes. 
$S$ is not expected to be appreciable in the metallic state of these materials due to large charge carrier density, n  and large density of states at the E$_F$, $N(E_F)$. The predicted small values of S in R\sku compounds has been confirmed experimentally ~\cite{thermoelectric_capabilities,thermoelectric_capabilities1,GumeniukJP11,MaplePRB14}. Although S is small in the stoichiometric compounds, doping La\sku with electrons, by replacing Ge with Sb, namely LaPt$_4$Ge$_{12-x}$Sb$_x$, drives the system close to a metal-insulator transition, and beyond $x$ = 5 the system is insulating. Close to the metal-insulator transition, the thermoelectric properties of the  LaPt$_4$Ge$_{12-x}$Sb$_x$ compound are greatly enhanced vs. the undoped La\sku compound~\cite{thermoelectric_capabilities}. Experiments showed that increasing the electron doping enhances the overall temperature dependence of the resistivity $\rho(T)=1/\sigma$ and the charge carrier density $n$, both consistent with an overall enhanced temperature dependence of S~\cite{thermoelectric_capabilities}. To understand these experimental findings, in Ref.~\onlinecite{thermoelectric_capabilities}, the temperature dependence of the Seebeck coefficient S(T) was computed for various $x$ values and it was shown that S(T) increases drastically in the vicinity of the metal-insulator transition, e.g. by an order of magnitude at room temperature. For the La\sku, the possibility of the metal-insulator transition was connected with the existence of a pseudogap above $E_F$ for $x$=0.

Our total DOS calculations for R\sku (see Fig.~\ref{Figure3_GDOS}) show the existence of notable pseudogaps around 1-1.5 eV above $E_F$, independently of the theoretical scenario we used in our calculations (PBE, PBE+SOC or PBE+SOC+U).  Based on Eq.~(\ref{eq:S}) for $S(T)$, the presence of such a pseudogap suggests that large values of S could be produced by tuning these metallic systems towards an insulating state or to a "bad metal" state by electron doping.  In addition, the measurements of the temperature dependent resistivity in PrPt$_4$Ge$_{12-x}$Sb$_x$ ~\cite{MaplePRB16} which show the same trend as in LaPt$_4$Ge$_{12-x}$Sb$_x$ ~\cite{thermoelectric_capabilities} further suggests the possibility of increasing ZT in R\sku. 

Based on our total DOS calculations showing pseudogap features above the E$_F$ and using the similarity of previous resistivity measurements in R\sku, we suggest that electron doped Ce\sku and Pr\sku compounds could also exhibit larger figures of merit than the stoichiometric compounds,  as it was already proved for doped La\sku.

\section{Discussion}
\label{sec_Discussion}

We have presented a comprehensive study of band structure, DOS and FS for f-electron systems 
of type R\sku (where R = La, Ce, Pr), using various theoretical scenarios such as PBE, PBE+SOC and PBE+SOC+U within the two choices of treating the $f$-electrons, as valence or core electrons. Our calculations show several bands crossing the E$_F$ giving rise to multiple FS. 

As expected, when the $f$-electrons are treated as part of core, the three compound show almost identical band structure, and thus similar FS, for a given theoretical scenario. Scenarios where SOC is present or absent, show different FS due to the fact that SOC lifts band degeneracy and shift bands, thus increasing the number of FS (especially at $\Gamma$ point) and changing slightly the FS topology.

When the $f$-electrons are treated in valence the three compounds show some slight differences. Although SOC has the same trends on the band structure as mentioned before, the increasing number of $f$-electrons from 0 to 2 and the increase of correlations by adding the Hubbard parameter U on top of the PBE+SOC approximation, makes things more interesting. Our calculations within PBE+SOC approximation reveal the presence of six FS for the La\sku, four of them having a hole like character and the other two electron like character. While increasing the number of $f$-electrons, the hole pockets get filled and three of them disappear going from La\sku to Pr\sku, while the electron pockets increase their surface. The effects of filling the hole pockets at $\Gamma$ point are reversed with increasing correlations effects between the $f$-electrons within the PBE+SOC+U approximation. Although these compounds have two electron-like and one hole-like FS in common, independently of the theoretical scenario used, the hole pockets at $\Gamma$ point  are dependent on the theoretical scenario we used in our calculations. Thus, de Hass van Alphen~\cite{Haas_van_ROURKE2012324,Haas_van_PhysRevB.91.121105,Haas_van_PhysRevB.88.155138} experiments are desirable to validate the topology and the number of FS in R\sku compounds. 

While our theoretical calculations cannot directly explain superconductivity, they may have significance with respect to the multi-band superconductivity or anisotropic nature of the pairing, as  indicated by experiments on Pr\sku and La\sku. Our calculations find a small amount of Pr-4$f$ states admixed with Ge-$p$ states in Pr\sku, a feature obviously different from that in La\sku. This is confirmed by soft X-ray photoemission experiments on Pr\sku~\cite{NakamuraJPSJ10,NakamuraPRB12} that show a finite contribution of Pr 4$f$ to the states near $E_F$.  Additionally, superconducting gaps are effected~\cite{SinghPRB16,MaplePRB16} by Ce substitution of Pr, i.e. Pr$_{1-x}$Ce$_x$\sku, suggesting that the filler atoms may affect superconductivity. Based on these, it is conceivable that superconductivity is a consequence of Cooper pairs formed by the s, p and d electrons of the Pt and Ge atoms in the La\sku, while superconductivity in the Pr\sku is more unusual since in addition to s, p, d electrons we also have $f$-electrons at the E$_F$.  As we have remarked earlier, since the Pr\sku has  anisotropic $f$ character on the FS, while the La\sku does not, this may lead to different electron-phonon couplings, and hence different forms of pairing. Thus it would be interesting to calculate electron-phonon couplings in these systems.

In addition, we have shown that the total DOS has a pseudogap which could imply the tunability of these materials through a metal to insulator transition and consequently the possibility of improving the thermoelectric properties of these materials. While DFT calculations are not able to treat the energy variation of mobility (and hence lifetime) that can modify results for $S(T)$,
a method such as DMFT, which is able to calculate finite-T lifetime, may provide further understanding of $S(T)$.

Finally we mention that our results are consistent with previous DOS calculations~\cite{GumeniukPRL08, GumeniukJP11} that comprise a limited set of results and a small subset of our more extensive work. However, our band structure and FS results differ substantially from those presented in a recent paper~\cite{ChandraPhilmag16}, which used the PBEsol approximation. We also performed calculations using the PBEsol approximation and found the results to be very similar to those obtained using the PBE approximation. Our results also reveal challenges confronted by DFT methods and its extension for correlated materials. We showed that the DFT +U method can only qualitatively explain the soft X-ray photoemission experiments on Pr\sku. Thus, it would be desirable to use more advance methods for correlated materials, such as DFT + eDMFT, in order to explain quantitatively the experimental results.

\section{Acknowledgements}
\label{sec_Acknowledgements}

MW acknowledges support from the Department of Energy under grant DE-SC0014506. KQ  acknowledges a QuantEmX grant from ICAM and the Gordon and Betty Moore Foundation, Grant GBMF5305, which partly funded this work. G.L.P. and K.H. were supported by the  U.S. Department of Energy, Office of Science, Basic Energy Sciences, as a part of the Computational Materials Science Program, funded by the  U.S. Department of Energy, Office of Science, Basic Energy Sciences, Materials Sciences and Engineering Division. We thank Di Xiao, G. Malcolm Stocks, Gabriel Kotliar, Greg Boebinger, and Carmen Almasan for useful discussions.

\section{Appendix}
\label{sec_Appendix}

\subsection{Electronic band structure: $f$-electrons in 'Core'}

\begin{figure*}[!htb]
\includegraphics[width=0.85\linewidth]{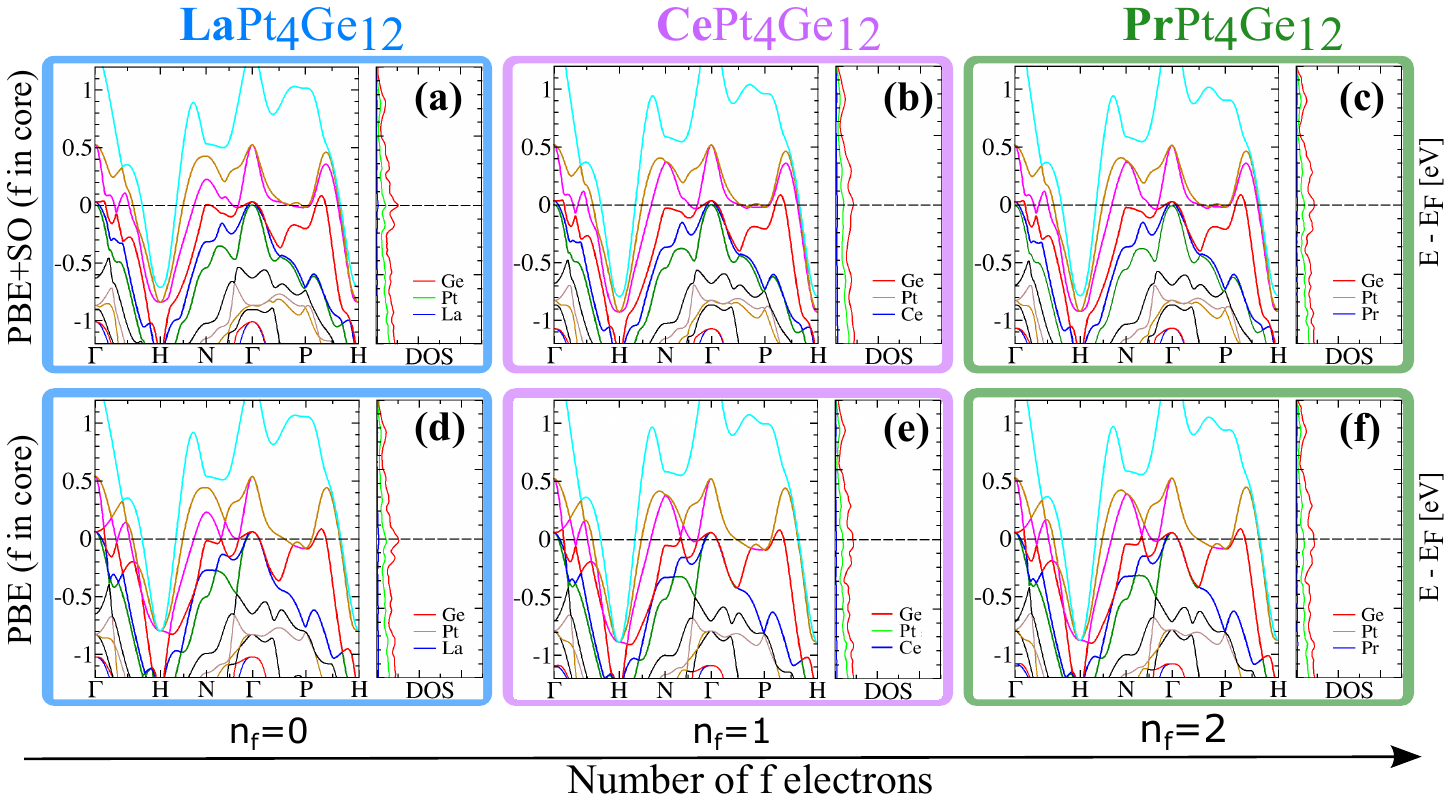}
\caption{(Color online) \textbf{Electronic Band Structure and Atom-Projected DOS} results for R\sku compounds obtained using the VASP code. These calculations were performed by placing the $f$ electrons into core. Each column corresponds to a particular compounds whose chemical formula is printed at the top of the column. Each row corresponds to results from calculations using a particular approximation who's label is printed at the beginning of the row. Each panel presents results for a given compound within a given approximation. \textit{On the left side of each panel} we show the electronic band structure (energy in eV on the vertical axis versus momentum on the horizontal axis). Fermi energy is marked by the horizontal dashed line. \textit{On the right side of each panel} we show atom-projected DOS (each unit on the DOS axis represents 1 state/eV/f.u.).} 
\label{Figure2_BANDS_Appendix}
\end{figure*}

In some systems, the states of $f$-electrons behave as inert localized states, hybridizing very little or at all with the other states in the system. To understand the effects of $f$-electron hybridization, we present the results of calculations when the $f$-electrons are treated as core electrons. In Fig.~\ref{Figure2_BANDS_Appendix} we show these calculations within the PBE and PBE+SOC approximations. We find that the electronic band structure within a given approximation for the R\sku compounds is almost identical when the $f$-electrons are treated as core. This implies that all the interesting low temperature properties of Ce\sku and Pr\sku are mostly due to the $f$ electrons states and their hybridization with the other states in the system.

\subsection{Fat-band Representation}

\begin{figure*}[!htb]
\includegraphics[width=0.85\linewidth]{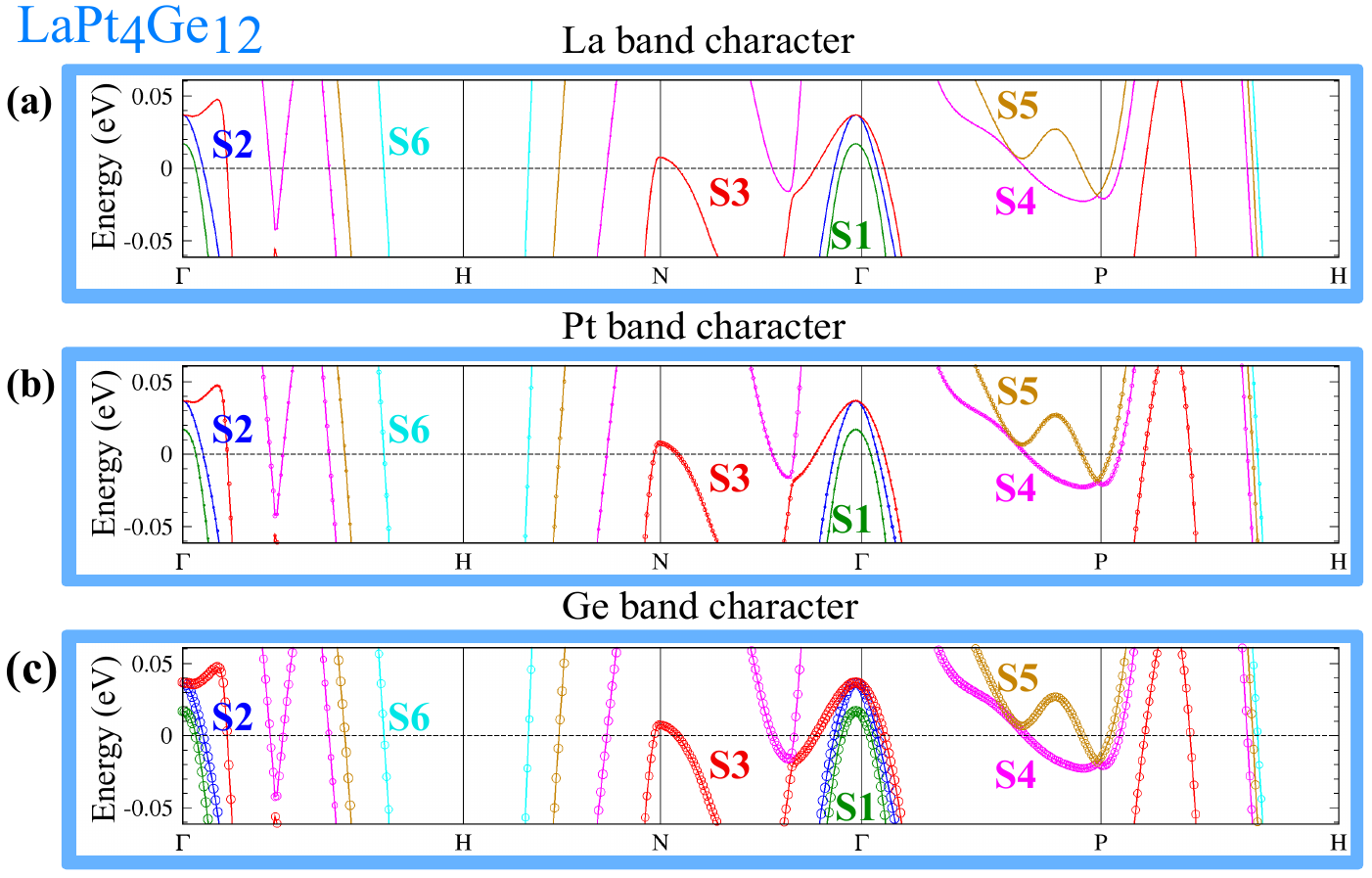}
\caption{(Color online) \textbf{Electronic band structure of LaPt$_4$Ge$_{12}$ showing the atomic character of the bands}. This figure enlarges the energy range near the Fermi energy of the band structure previously shown in Fig.~\ref{Figure4_FS}. The sizes of the open circles at a given energy and momentum represent the wave function amplitude corresponding to a given atom as labeled in each panel.}
\label{Figure6_BCP}.
\end{figure*}
\begin{figure*}[!htb]
\includegraphics[width=0.85\linewidth]{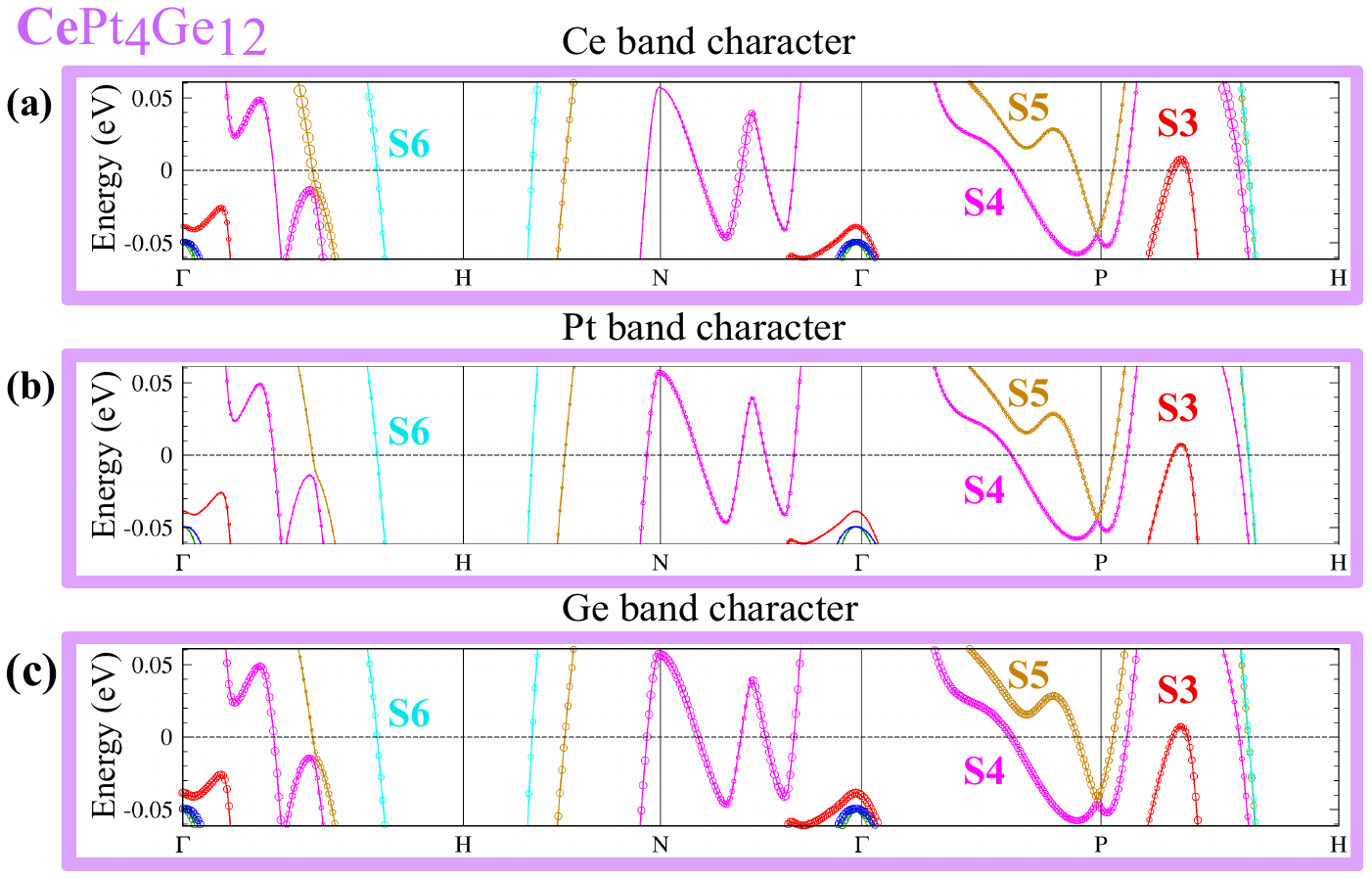}
\caption{(Color online) \textbf{Electronic band structure of CePt$_4$Ge$_{12}$ showing the atomic character of the bands}. This figure enlarges the energy range near the Fermi energy of the band structure previously shown in Fig.~\ref{Figure4_FS}. The sizes of the open circles at a given energy and momentum represent the wave function amplitude corresponding to a given atom as labeled in each panel.}
\label{Figure7_BCP}
\end{figure*}
\begin{figure*}[!htb]
\includegraphics[width=0.85\linewidth]{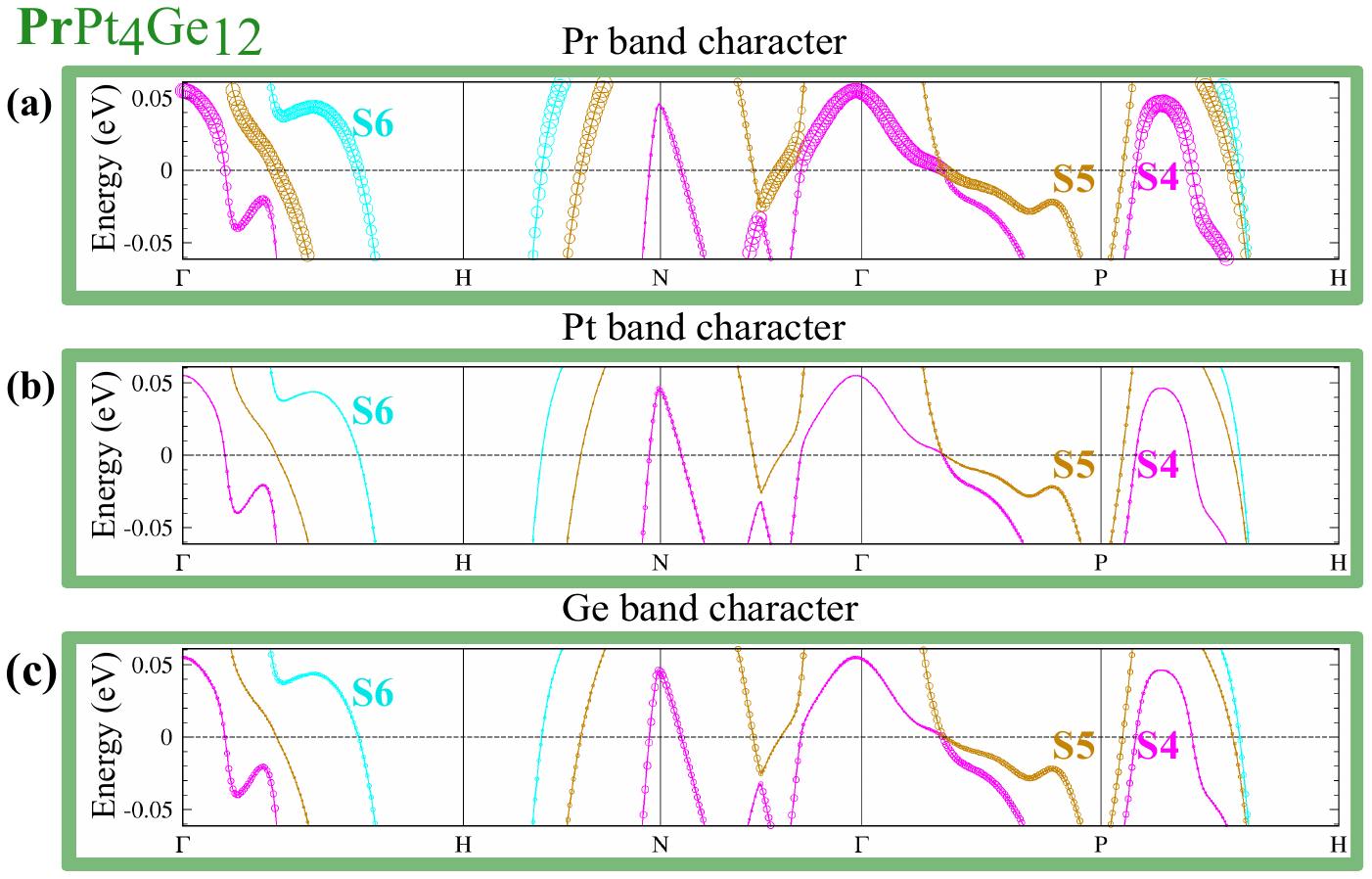}
\caption{(Color online) \textbf{Electronic band structure of PrPt$_4$Ge$_{12}$ showing the atomic character of the bands}. This figure enlarges the energy range near the Fermi energy of the band structure previously shown in Fig.~\ref{Figure4_FS}. The sizes of the open circles at a given energy and momentum represent the wave function amplitude corresponding to a given atom as labeled in each panel.}
\label{Figure8_BCP}
\end{figure*}

By projecting the band wave-function onto the atomic orbitals of a particular atom, we obtain the band structures shown by solid lines + open circles, where the size of the circles at an energy and momentum point are proportional to the absolute value of the wave-function amplitude corresponding to that given atom~\cite{FatBands}.  It may be noted that in these plots the size of the circle corresponds to the atomic character and is shown for one atom only, even thought the band wave-functions forming the bands have contributions from all the atoms (of the same type) within the unit cell.

In Fig.~\ref{Figure6_BCP}, we plot the electronic band structure with the atomic character for La\sku. Looking at the (a), (b) and (c) panels corresponding to the La, Pt and Ge atomic character, we see that around the Fermi energy the electronic character comes mostly from the Ge and Pt atoms, consistent with the previously presented atom projected DOS in Fig.~2. At the same time, looking at the the first column in Figure~\ref{Figure3_GDOS}, we see that for Pt atoms the largest DOS at the E$_F$ comes $p$ and $d$ states and for the case of Ge ion the maximum DOS comes from $p$ states. Thus, it may tempting to say that for the La\sku compound the electrons involved in superconductivity would mostly be $p$ Ge states plus some $p$ and $d$ Pt states. With reference to the notation in Fig. 4 for FS, here we can see that the S1, S2 and S6 FS are mostly Ge in character, while S3, S4 and S5 have contributions from both Ge and Pt ions. Notably, La is largely absent from the FS.

Figure~\ref{Figure7_BCP} shows fat-band plots for Ce\sku. The discussion here follows that of Fig.~\ref{Figure6_BCP} for La\sku, with the additional fact that for Ce\sku the FS have moderate $f$ character in addition.

Likewise, in Fig.~\ref{Figure8_BCP}, we show fat-band plots for Pr\sku. A similar discussion can be made as before for the case of La\sku and Ce\sku with the additional fact that for Pr\sku the FS have strong $f$ character. In addition, for the Pr\sku compound if we look at Fig.~\ref{Figure8_BCP}(a), where we show the Pr contribution to the FS, we see that the $f$ character of  the FS is anisotropic. For example, compare the strength (size of the circles) of Pr $f$ character at the two intersections of the S4 band with the E$_F$ between points P and N. In analogy with Pb, one could suggest that anisotropic band character could lead to anisotropic electron-phonon couplings for these FS. This makes for a strong case for calculating electron-phonon couplings in the La\sku and Pr\sku skutterudites, and exploring if $f$-electron coupling leads to multi-gap pairing in the  case of Pr\sku.

\subsection{Effects of $U$ on Bands and Fermi Surfaces}

\begin{figure*}
\includegraphics[width=0.85\linewidth]{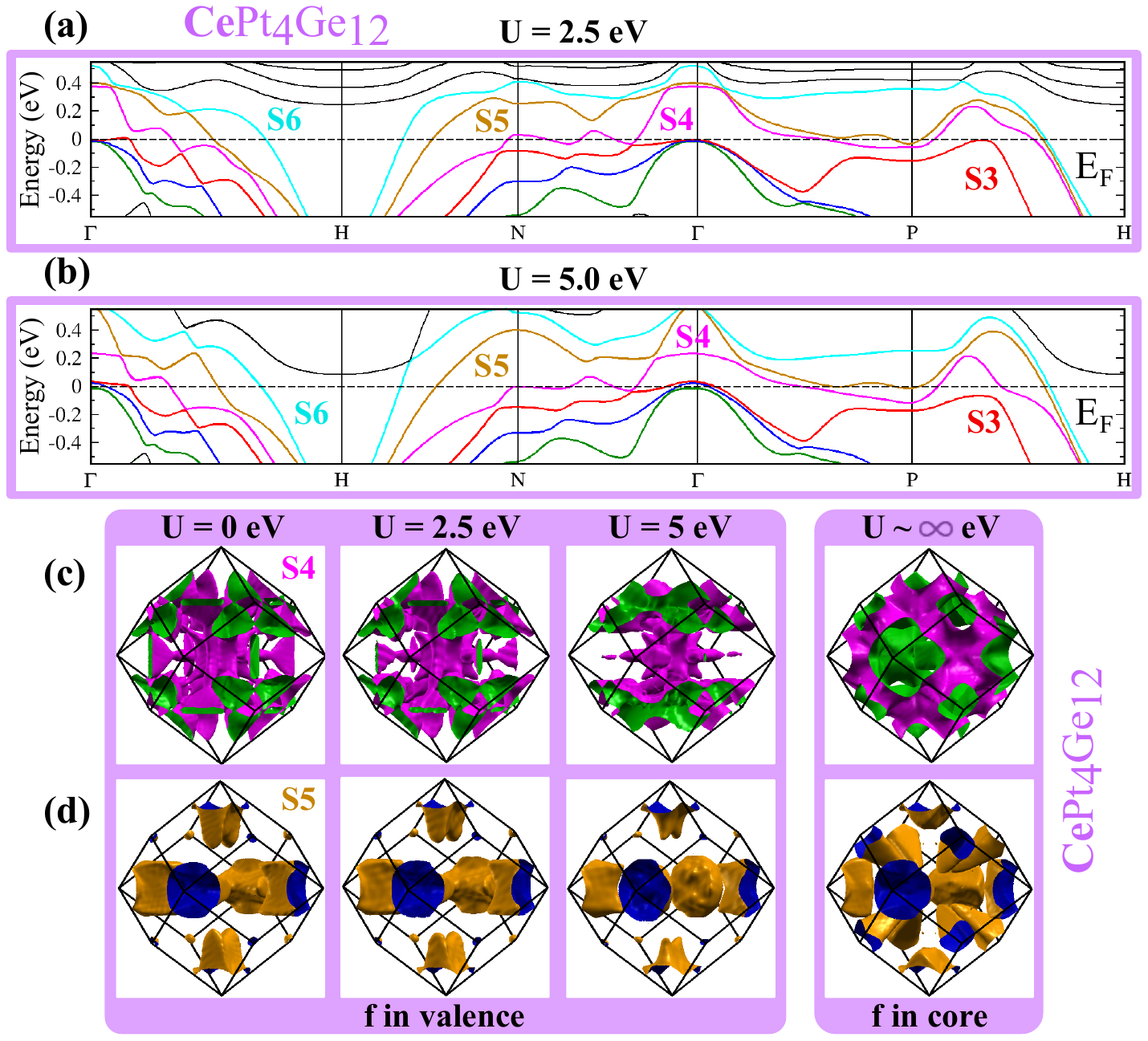}
\caption{(Color online) \textbf{Correlations effects on the electronic band structure and FS for CePt$_4$Ge$_{12}$}. The results presented in this figure were obtained using the Wien2k code with the PBE+SOC+U approximation treating the $f$-electrons as valence electrons (exception are the results presented in the U $\sim\infty$ eV panel,which correspond to calculations using the  PBE+SOC approximation treating the $f$-electrons as core electrons). The electronic band structure plots shown here are on the same scale, plotted with the same color code for the lines representing the bands and with the same labels for the Fermi surfaces as in panel (d) of Fig.~~\ref{Figure4_FS}. Panels (a) and (b) show the electronic band structure for calculations with U of 2.5 and 5 eV. Panels (c) and (d) show the evolution with correlations for two distinct FS.} 
\label{Ce-bands-U}
\end{figure*}

\begin{figure*}
\includegraphics[width=0.85\linewidth]{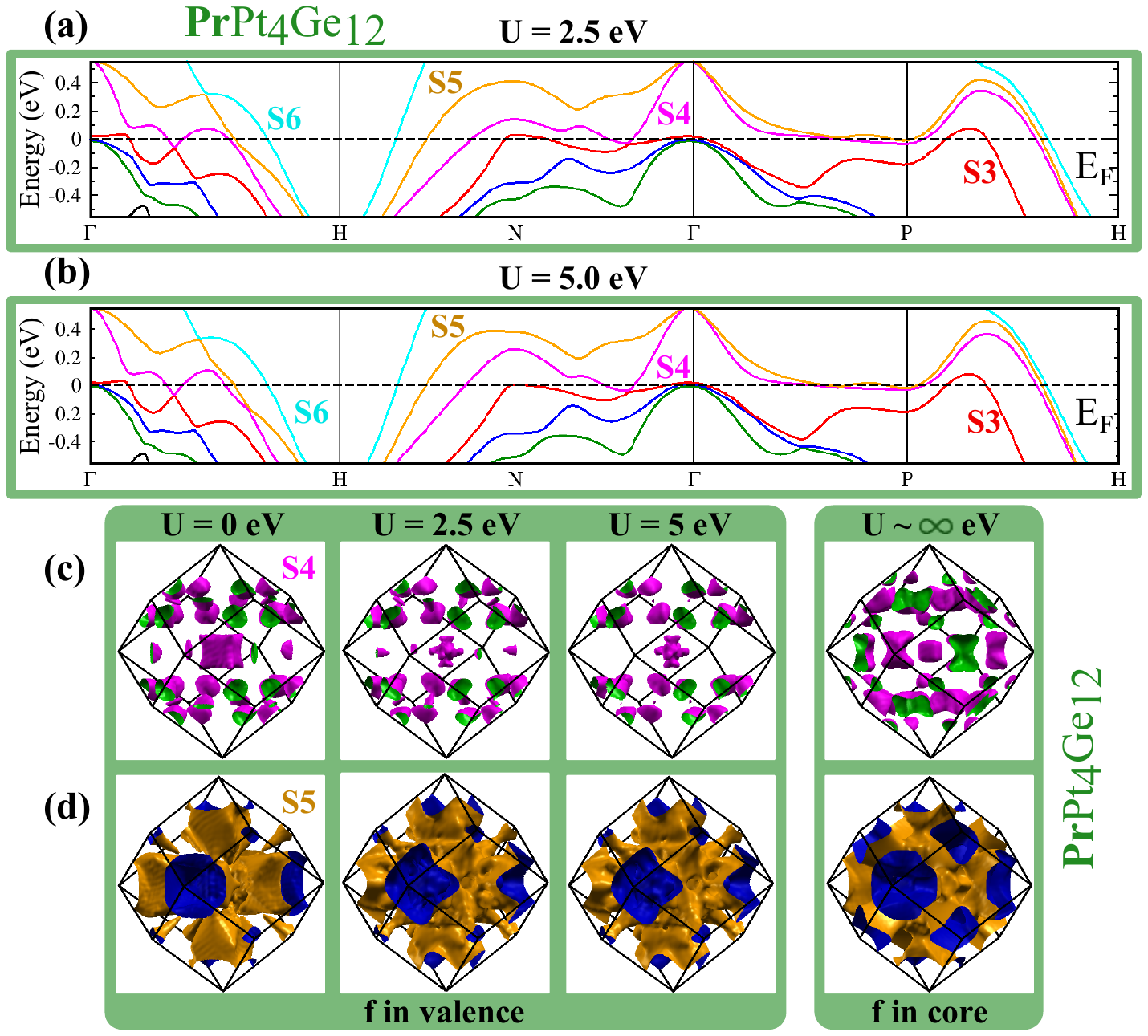}
\caption{(Color online) \textbf{Correlations effects on the electronic band structure and FS for PrPt$_4$Ge$_{12}$}. The results presented in this figure were obtained using the Wien2k code with the PBE+SOC+U approximation treating the $f$-electrons as valence electrons (exception are the results presented in the U $\sim\infty$ eV panel,which correspond to calculations using the  PBE+SO approximation treating the $f$-electrons as core electrons). The electronic band structure plots shown here are on the same scale, plotted with the same color code for the lines representing the bands and with the same labels for the Fermi surfaces as in panel (f) of Fig.~~\ref{Figure4_FS}. Panels (a) and (b) show the electronic band structure for calculations with U of 2.5 and 5 eV. Panels (c) and (d) show the evolution with correlations for two distinct FS.} 
\label{Pr-bands-U}
\end{figure*}

From band structure plots, Fig.~\ref{Ce-bands-U} (a) - (b), and Fig. ~\ref{Pr-bands-U} (a) - (b),  it is obvious that the band corresponding to the S6 FS doesn't change much as we increase the correlation, thus this FS is quite robust which indicates that it is probably mostly $s$, $p$ and $d$ in character. On the other hand, plotting the S4 and S5 FS versus U (see Fig.~\ref{Ce-bands-U} (c) - (d), and Fig. ~\ref{Pr-bands-U} (c) - (d), we find that S4 is the most affected, followed by S5. This suggests that the S4 and S5 FS have stronger $f$ character than S6. To confirm the contribution of $f$ states to the S4 and S5 FS for $U~\ge~0~eV$, we have also computed the FS for the case where the $f$-electrons are treated as core electrons (in this case the FS has no $f$ contribution). These results are shown in  Fig.~\ref{Ce-bands-U} (c) - (d), and Fig. ~\ref{Pr-bands-U} (c) - (d), labeled as $U~\sim~\infty~eV$. The fact that the topology of the FS for $U~\ge~0~eV$ is different than that of $U~\sim~\infty~eV$, confirms once again the presence of $f$ character in the FS of Ce\sku and Pr\sku. In addition, if we compare the bands in Fig.~\ref{Figure4_FS}{d} ($U$=0) for Ce\sku with Fig.~\ref{Ce-bands-U} ($U>0$)  we see for example, that  with increasing $U$, band S3 falls below $E_F$ along the path P-H, but rises above at $\Gamma$.

Comparing Fig.~\ref{Figure4_FS}(f) ($U$=0) for Pr\sku with Fig.~\ref{Pr-bands-U} ($U>0$), we see that S3 similarly rises above $E_F$ at $\Gamma$, creating a new FS sheet. Thus, in both Ce\sku and Pr\sku compounds, we find that depending on the strength of correlation parameter U, the S1, S2 and S3 FS can be present or absent. As mentioned in the discussion section, de Haas-van Alphen measurements could sort out the puzzle of FS number and indirectly give information of the correlation strength in the Ce\sku and Pr\sku compounds.

\bibliography{Bib_Skutterudites_PWHQ}

\begin{thebibliography}{74}
\expandafter\ifx\csname natexlab\endcsname\relax\def\natexlab#1{#1}\fi
\expandafter\ifx\csname bibnamefont\endcsname\relax
  \def\bibnamefont#1{#1}\fi
\expandafter\ifx\csname bibfnamefont\endcsname\relax
  \def\bibfnamefont#1{#1}\fi
\expandafter\ifx\csname citenamefont\endcsname\relax
  \def\citenamefont#1{#1}\fi
\expandafter\ifx\csname url\endcsname\relax
  \def\url#1{\texttt{#1}}\fi
\expandafter\ifx\csname urlprefix\endcsname\relax\def\urlprefix{URL }\fi
\providecommand{\bibinfo}[2]{#2}
\providecommand{\eprint}[2][]{\url{#2}}

\bibitem[{\citenamefont{Jeitschko and Braun}(1977)}]{CS_Jeitschko}
\bibinfo{author}{\bibfnamefont{W.}~\bibnamefont{Jeitschko}} \bibnamefont{and}
  \bibinfo{author}{\bibfnamefont{D.}~\bibnamefont{Braun}},
  \bibinfo{journal}{Acta Crystallographica Section B}
  \textbf{\bibinfo{volume}{33}}, \bibinfo{pages}{3401} (\bibinfo{year}{1977}),
  \urlprefix\url{https://doi.org/10.1107/S056774087701108X}.

\bibitem[{\citenamefont{Gumeniuk
  et~al.}(2010{\natexlab{a}})\citenamefont{Gumeniuk, Borrmann, Ormeci, Rosner,
  Schnelle, Nicklas, Grin, and Leithe-Jasper}}]{R_Gumeniuk_CS_2010}
\bibinfo{author}{\bibfnamefont{R.}~\bibnamefont{Gumeniuk}},
  \bibinfo{author}{\bibfnamefont{H.}~\bibnamefont{Borrmann}},
  \bibinfo{author}{\bibfnamefont{A.}~\bibnamefont{Ormeci}},
  \bibinfo{author}{\bibfnamefont{H.}~\bibnamefont{Rosner}},
  \bibinfo{author}{\bibfnamefont{W.}~\bibnamefont{Schnelle}},
  \bibinfo{author}{\bibfnamefont{M.}~\bibnamefont{Nicklas}},
  \bibinfo{author}{\bibfnamefont{Y.}~\bibnamefont{Grin}}, \bibnamefont{and}
  \bibinfo{author}{\bibfnamefont{A.}~\bibnamefont{Leithe-Jasper}},
  \bibinfo{journal}{Zeitschrift f$\ddot{u}$r Kristallographie Crystalline
  Materials} \textbf{\bibinfo{volume}{225}}, \bibinfo{pages}{531}
  (\bibinfo{year}{2010}{\natexlab{a}}).

\bibitem[{\citenamefont{B.~Maple et~al.}(2008)\citenamefont{B.~Maple, Henkie,
  E.~Baumbach, A.~Sayles, P.~Butch, Ho, Yanagisawa, M.~Yuhasz, Wawryk, Cichorek
  et~al.}}]{BCS_KL_Maple}
\bibinfo{author}{\bibfnamefont{M.}~\bibnamefont{B.~Maple}},
  \bibinfo{author}{\bibfnamefont{Z.}~\bibnamefont{Henkie}},
  \bibinfo{author}{\bibfnamefont{R.}~\bibnamefont{E.~Baumbach}},
  \bibinfo{author}{\bibfnamefont{T.}~\bibnamefont{A.~Sayles}},
  \bibinfo{author}{\bibfnamefont{N.}~\bibnamefont{P.~Butch}},
  \bibinfo{author}{\bibfnamefont{P.-C.} \bibnamefont{Ho}},
  \bibinfo{author}{\bibfnamefont{T.}~\bibnamefont{Yanagisawa}},
  \bibinfo{author}{\bibfnamefont{W.}~\bibnamefont{M.~Yuhasz}},
  \bibinfo{author}{\bibfnamefont{R.}~\bibnamefont{Wawryk}},
  \bibinfo{author}{\bibfnamefont{T.}~\bibnamefont{Cichorek}},
  \bibnamefont{et~al.}, \bibinfo{journal}{Journal of the Physical Society of
  Japan} \textbf{\bibinfo{volume}{77}}, \bibinfo{pages}{7}
  (\bibinfo{year}{2008}), \eprint{https://doi.org/10.1143/JPSJS.77SA.7},
  \urlprefix\url{https://doi.org/10.1143/JPSJS.77SA.7}.

\bibitem[{\citenamefont{Aoki et~al.}(2007)\citenamefont{Aoki, Tayama,
  Sakakibara, Kuwahara, Iwasa, Kohgi, Higemoto, E.~MacLaughlin, Sugawara, and
  Sato}}]{GLP5_Unconventional_Superconductivity}
\bibinfo{author}{\bibfnamefont{Y.}~\bibnamefont{Aoki}},
  \bibinfo{author}{\bibfnamefont{T.}~\bibnamefont{Tayama}},
  \bibinfo{author}{\bibfnamefont{T.}~\bibnamefont{Sakakibara}},
  \bibinfo{author}{\bibfnamefont{K.}~\bibnamefont{Kuwahara}},
  \bibinfo{author}{\bibfnamefont{K.}~\bibnamefont{Iwasa}},
  \bibinfo{author}{\bibfnamefont{M.}~\bibnamefont{Kohgi}},
  \bibinfo{author}{\bibfnamefont{W.}~\bibnamefont{Higemoto}},
  \bibinfo{author}{\bibfnamefont{D.}~\bibnamefont{E.~MacLaughlin}},
  \bibinfo{author}{\bibfnamefont{H.}~\bibnamefont{Sugawara}}, \bibnamefont{and}
  \bibinfo{author}{\bibfnamefont{H.}~\bibnamefont{Sato}},
  \bibinfo{journal}{Journal of the Physical Society of Japan}
  \textbf{\bibinfo{volume}{76}}, \bibinfo{pages}{051006}
  (\bibinfo{year}{2007}), \eprint{https://doi.org/10.1143/JPSJ.76.051006},
  \urlprefix\url{https://doi.org/10.1143/JPSJ.76.051006}.

\bibitem[{\citenamefont{Maisuradze et~al.}(2009)\citenamefont{Maisuradze,
  Nicklas, Gumeniuk, Baines, Schnelle, Rosner, Leithe-Jasper, Grin, and
  Khasanov}}]{MaisuradzePRL09}
\bibinfo{author}{\bibfnamefont{A.}~\bibnamefont{Maisuradze}},
  \bibinfo{author}{\bibfnamefont{M.}~\bibnamefont{Nicklas}},
  \bibinfo{author}{\bibfnamefont{R.}~\bibnamefont{Gumeniuk}},
  \bibinfo{author}{\bibfnamefont{C.}~\bibnamefont{Baines}},
  \bibinfo{author}{\bibfnamefont{W.}~\bibnamefont{Schnelle}},
  \bibinfo{author}{\bibfnamefont{H.}~\bibnamefont{Rosner}},
  \bibinfo{author}{\bibfnamefont{A.}~\bibnamefont{Leithe-Jasper}},
  \bibinfo{author}{\bibfnamefont{Y.}~\bibnamefont{Grin}}, \bibnamefont{and}
  \bibinfo{author}{\bibfnamefont{R.}~\bibnamefont{Khasanov}},
  \bibinfo{journal}{Phys. Rev. Lett.} \textbf{\bibinfo{volume}{103}},
  \bibinfo{pages}{147002} (\bibinfo{year}{2009}),
  \urlprefix\url{https://link.aps.org/doi/10.1103/PhysRevLett.103.147002}.

\bibitem[{\citenamefont{Maisuradze et~al.}(2010)\citenamefont{Maisuradze,
  Schnelle, Khasanov, Gumeniuk, Nicklas, Rosner, Leithe-Jasper, Grin, Amato,
  and Thalmeier}}]{UncSuper_PhysRevB.82.024524}
\bibinfo{author}{\bibfnamefont{A.}~\bibnamefont{Maisuradze}},
  \bibinfo{author}{\bibfnamefont{W.}~\bibnamefont{Schnelle}},
  \bibinfo{author}{\bibfnamefont{R.}~\bibnamefont{Khasanov}},
  \bibinfo{author}{\bibfnamefont{R.}~\bibnamefont{Gumeniuk}},
  \bibinfo{author}{\bibfnamefont{M.}~\bibnamefont{Nicklas}},
  \bibinfo{author}{\bibfnamefont{H.}~\bibnamefont{Rosner}},
  \bibinfo{author}{\bibfnamefont{A.}~\bibnamefont{Leithe-Jasper}},
  \bibinfo{author}{\bibfnamefont{Y.}~\bibnamefont{Grin}},
  \bibinfo{author}{\bibfnamefont{A.}~\bibnamefont{Amato}}, \bibnamefont{and}
  \bibinfo{author}{\bibfnamefont{P.}~\bibnamefont{Thalmeier}},
  \bibinfo{journal}{Phys. Rev. B} \textbf{\bibinfo{volume}{82}},
  \bibinfo{pages}{024524} (\bibinfo{year}{2010}),
  \urlprefix\url{https://link.aps.org/doi/10.1103/PhysRevB.82.024524}.

\bibitem[{\citenamefont{Takeda and Ishikawa}(2000)}]{non_fermi_liquid1}
\bibinfo{author}{\bibfnamefont{N.}~\bibnamefont{Takeda}} \bibnamefont{and}
  \bibinfo{author}{\bibfnamefont{M.}~\bibnamefont{Ishikawa}},
  \bibinfo{journal}{Journal of the Physical Society of Japan}
  \textbf{\bibinfo{volume}{69}}, \bibinfo{pages}{868} (\bibinfo{year}{2000}),
  \eprint{https://doi.org/10.1143/JPSJ.69.868},
  \urlprefix\url{https://doi.org/10.1143/JPSJ.69.868}.

\bibitem[{\citenamefont{Yamamoto et~al.}(2006)\citenamefont{Yamamoto, Wada,
  Shirotani, and Sekine}}]{non_fermi_liquid2}
\bibinfo{author}{\bibfnamefont{A.}~\bibnamefont{Yamamoto}},
  \bibinfo{author}{\bibfnamefont{S.}~\bibnamefont{Wada}},
  \bibinfo{author}{\bibfnamefont{I.}~\bibnamefont{Shirotani}},
  \bibnamefont{and} \bibinfo{author}{\bibfnamefont{C.}~\bibnamefont{Sekine}},
  \bibinfo{journal}{Journal of the Physical Society of Japan}
  \textbf{\bibinfo{volume}{75}}, \bibinfo{pages}{063703}
  (\bibinfo{year}{2006}), \eprint{https://doi.org/10.1143/JPSJ.75.063703},
  \urlprefix\url{https://doi.org/10.1143/JPSJ.75.063703}.

\bibitem[{\citenamefont{Sekine et~al.}(1997)\citenamefont{Sekine, Uchiumi,
  Shirotani, and Yagi}}]{MIT_PhysRevLett.79.3218}
\bibinfo{author}{\bibfnamefont{C.}~\bibnamefont{Sekine}},
  \bibinfo{author}{\bibfnamefont{T.}~\bibnamefont{Uchiumi}},
  \bibinfo{author}{\bibfnamefont{I.}~\bibnamefont{Shirotani}},
  \bibnamefont{and} \bibinfo{author}{\bibfnamefont{T.}~\bibnamefont{Yagi}},
  \bibinfo{journal}{Phys. Rev. Lett.} \textbf{\bibinfo{volume}{79}},
  \bibinfo{pages}{3218} (\bibinfo{year}{1997}),
  \urlprefix\url{https://link.aps.org/doi/10.1103/PhysRevLett.79.3218}.

\bibitem[{\citenamefont{Yoshizawa et~al.}(2005)\citenamefont{Yoshizawa,
  Nakanishi, Oikawa, Sekine, Shirotani, R.~Saha, Sugawara, and
  Sato}}]{multipolar_ordering}
\bibinfo{author}{\bibfnamefont{M.}~\bibnamefont{Yoshizawa}},
  \bibinfo{author}{\bibfnamefont{Y.}~\bibnamefont{Nakanishi}},
  \bibinfo{author}{\bibfnamefont{M.}~\bibnamefont{Oikawa}},
  \bibinfo{author}{\bibfnamefont{C.}~\bibnamefont{Sekine}},
  \bibinfo{author}{\bibfnamefont{I.}~\bibnamefont{Shirotani}},
  \bibinfo{author}{\bibfnamefont{S.}~\bibnamefont{R.~Saha}},
  \bibinfo{author}{\bibfnamefont{H.}~\bibnamefont{Sugawara}}, \bibnamefont{and}
  \bibinfo{author}{\bibfnamefont{H.}~\bibnamefont{Sato}},
  \bibinfo{journal}{Journal of the Physical Society of Japan}
  \textbf{\bibinfo{volume}{74}}, \bibinfo{pages}{2141} (\bibinfo{year}{2005}),
  \eprint{https://doi.org/10.1143/JPSJ.74.2141},
  \urlprefix\url{https://doi.org/10.1143/JPSJ.74.2141}.

\bibitem[{\citenamefont{Yan et~al.}(2012)\citenamefont{Yan, M\"uchler, Qi,
  Zhang, and Felser}}]{TI1_PhysRevB.85.165125}
\bibinfo{author}{\bibfnamefont{B.}~\bibnamefont{Yan}},
  \bibinfo{author}{\bibfnamefont{L.}~\bibnamefont{M\"uchler}},
  \bibinfo{author}{\bibfnamefont{X.-L.} \bibnamefont{Qi}},
  \bibinfo{author}{\bibfnamefont{S.-C.} \bibnamefont{Zhang}}, \bibnamefont{and}
  \bibinfo{author}{\bibfnamefont{C.}~\bibnamefont{Felser}},
  \bibinfo{journal}{Phys. Rev. B} \textbf{\bibinfo{volume}{85}},
  \bibinfo{pages}{165125} (\bibinfo{year}{2012}),
  \urlprefix\url{https://link.aps.org/doi/10.1103/PhysRevB.85.165125}.

\bibitem[{\citenamefont{Smith et~al.}(2011)\citenamefont{Smith, Banerjee,
  Pardo, and Pickett}}]{TI2_PhysRevLett.106.056401}
\bibinfo{author}{\bibfnamefont{J.~C.} \bibnamefont{Smith}},
  \bibinfo{author}{\bibfnamefont{S.}~\bibnamefont{Banerjee}},
  \bibinfo{author}{\bibfnamefont{V.}~\bibnamefont{Pardo}}, \bibnamefont{and}
  \bibinfo{author}{\bibfnamefont{W.~E.} \bibnamefont{Pickett}},
  \bibinfo{journal}{Phys. Rev. Lett.} \textbf{\bibinfo{volume}{106}},
  \bibinfo{pages}{056401} (\bibinfo{year}{2011}),
  \urlprefix\url{https://link.aps.org/doi/10.1103/PhysRevLett.106.056401}.

\bibitem[{\citenamefont{Masatoshi et~al.}(2008)\citenamefont{Masatoshi,
  Hitoshi, Ko-ichi, Takahito, Kuniyuki, Yuji, and Hideyuki}}]{TodaJPS08}
\bibinfo{author}{\bibfnamefont{T.}~\bibnamefont{Masatoshi}},
  \bibinfo{author}{\bibfnamefont{S.}~\bibnamefont{Hitoshi}},
  \bibinfo{author}{\bibfnamefont{M.}~\bibnamefont{Ko-ichi}},
  \bibinfo{author}{\bibfnamefont{S.}~\bibnamefont{Takahito}},
  \bibinfo{author}{\bibfnamefont{K.}~\bibnamefont{Kuniyuki}},
  \bibinfo{author}{\bibfnamefont{A.}~\bibnamefont{Yuji}}, \bibnamefont{and}
  \bibinfo{author}{\bibfnamefont{S.}~\bibnamefont{Hideyuki}},
  \bibinfo{journal}{Journal of the Physical Society of Japan}
  \textbf{\bibinfo{volume}{77}}, \bibinfo{pages}{124702}
  (\bibinfo{year}{2008}), \eprint{https://doi.org/10.1143/JPSJ.77.124702},
  \urlprefix\url{https://doi.org/10.1143/JPSJ.77.124702}.

\bibitem[{\citenamefont{Bauer et~al.}(2002)\citenamefont{Bauer, Frederick, Ho,
  Zapf, and Maple}}]{HFermionB_Maple_PhysRevB.65.100506}
\bibinfo{author}{\bibfnamefont{E.~D.} \bibnamefont{Bauer}},
  \bibinfo{author}{\bibfnamefont{N.~A.} \bibnamefont{Frederick}},
  \bibinfo{author}{\bibfnamefont{P.-C.} \bibnamefont{Ho}},
  \bibinfo{author}{\bibfnamefont{V.~S.} \bibnamefont{Zapf}}, \bibnamefont{and}
  \bibinfo{author}{\bibfnamefont{M.~B.} \bibnamefont{Maple}},
  \bibinfo{journal}{Phys. Rev. B} \textbf{\bibinfo{volume}{65}},
  \bibinfo{pages}{100506} (\bibinfo{year}{2002}),
  \urlprefix\url{https://link.aps.org/doi/10.1103/PhysRevB.65.100506}.

\bibitem[{\citenamefont{B.~Maple et~al.}(2002)\citenamefont{B.~Maple, Ho,
  S.~Zapf, A.~Frederick, D.~Bauer, M.~Yuhasz, M.~Woodward, and
  W.~Lynn}}]{HFermionB_Maple}
\bibinfo{author}{\bibfnamefont{M.}~\bibnamefont{B.~Maple}},
  \bibinfo{author}{\bibfnamefont{P.-C.} \bibnamefont{Ho}},
  \bibinfo{author}{\bibfnamefont{V.}~\bibnamefont{S.~Zapf}},
  \bibinfo{author}{\bibfnamefont{N.}~\bibnamefont{A.~Frederick}},
  \bibinfo{author}{\bibfnamefont{E.}~\bibnamefont{D.~Bauer}},
  \bibinfo{author}{\bibfnamefont{W.}~\bibnamefont{M.~Yuhasz}},
  \bibinfo{author}{\bibfnamefont{F.}~\bibnamefont{M.~Woodward}},
  \bibnamefont{and} \bibinfo{author}{\bibfnamefont{J.}~\bibnamefont{W.~Lynn}},
  \bibinfo{journal}{Journal of the Physical Society of Japan}
  \textbf{\bibinfo{volume}{71}}, \bibinfo{pages}{23} (\bibinfo{year}{2002}),
  \eprint{https://doi.org/10.1143/JPSJS.71S.23},
  \urlprefix\url{https://doi.org/10.1143/JPSJS.71S.23}.

\bibitem[{\citenamefont{Maple et~al.}(2005)\citenamefont{Maple, Frederick, Ho,
  Yuhasz, Sayles, Butch, Jeffries, and Taylor}}]{HFermionB_Maple_MAPLE2005830}
\bibinfo{author}{\bibfnamefont{M.}~\bibnamefont{Maple}},
  \bibinfo{author}{\bibfnamefont{N.}~\bibnamefont{Frederick}},
  \bibinfo{author}{\bibfnamefont{P.-C.} \bibnamefont{Ho}},
  \bibinfo{author}{\bibfnamefont{W.}~\bibnamefont{Yuhasz}},
  \bibinfo{author}{\bibfnamefont{T.}~\bibnamefont{Sayles}},
  \bibinfo{author}{\bibfnamefont{N.}~\bibnamefont{Butch}},
  \bibinfo{author}{\bibfnamefont{J.}~\bibnamefont{Jeffries}}, \bibnamefont{and}
  \bibinfo{author}{\bibfnamefont{B.}~\bibnamefont{Taylor}},
  \bibinfo{journal}{Physica B: Condensed Matter}
  \textbf{\bibinfo{volume}{359-361}}, \bibinfo{pages}{830 }
  (\bibinfo{year}{2005}), ISSN \bibinfo{issn}{0921-4526},
  \bibinfo{note}{proceedings of the International Conference on Strongly
  Correlated Electron Systems},
  \urlprefix\url{http://www.sciencedirect.com/science/article/pii/S0921452605002656}.

\bibitem[{\citenamefont{Maple et~al.}(2007)\citenamefont{Maple, Henkie, Yuhasz,
  Ho, Yanagisawa, Sayles, Butch, Jeffries, and
  Pietraszko}}]{MAGStates_MAPLE2007182}
\bibinfo{author}{\bibfnamefont{M.}~\bibnamefont{Maple}},
  \bibinfo{author}{\bibfnamefont{Z.}~\bibnamefont{Henkie}},
  \bibinfo{author}{\bibfnamefont{W.}~\bibnamefont{Yuhasz}},
  \bibinfo{author}{\bibfnamefont{P.-C.} \bibnamefont{Ho}},
  \bibinfo{author}{\bibfnamefont{T.}~\bibnamefont{Yanagisawa}},
  \bibinfo{author}{\bibfnamefont{T.}~\bibnamefont{Sayles}},
  \bibinfo{author}{\bibfnamefont{N.}~\bibnamefont{Butch}},
  \bibinfo{author}{\bibfnamefont{J.}~\bibnamefont{Jeffries}}, \bibnamefont{and}
  \bibinfo{author}{\bibfnamefont{A.}~\bibnamefont{Pietraszko}},
  \bibinfo{journal}{Journal of Magnetism and Magnetic Materials}
  \textbf{\bibinfo{volume}{310}}, \bibinfo{pages}{182 } (\bibinfo{year}{2007}),
  ISSN \bibinfo{issn}{0304-8853}, \bibinfo{note}{proceedings of the 17th
  International Conference on Magnetism},
  \urlprefix\url{http://www.sciencedirect.com/science/article/pii/S0304885306011516}.

\bibitem[{\citenamefont{Sato et~al.}(2009)\citenamefont{Sato, Aoki, Kikuchi,
  Sugawara, Higemoto, Ohishi, Ito, Heffner, Saha, Koda
  et~al.}}]{MAGStates_SATO2009749}
\bibinfo{author}{\bibfnamefont{H.}~\bibnamefont{Sato}},
  \bibinfo{author}{\bibfnamefont{Y.}~\bibnamefont{Aoki}},
  \bibinfo{author}{\bibfnamefont{D.}~\bibnamefont{Kikuchi}},
  \bibinfo{author}{\bibfnamefont{H.}~\bibnamefont{Sugawara}},
  \bibinfo{author}{\bibfnamefont{W.}~\bibnamefont{Higemoto}},
  \bibinfo{author}{\bibfnamefont{K.}~\bibnamefont{Ohishi}},
  \bibinfo{author}{\bibfnamefont{T.}~\bibnamefont{Ito}},
  \bibinfo{author}{\bibfnamefont{R.}~\bibnamefont{Heffner}},
  \bibinfo{author}{\bibfnamefont{S.}~\bibnamefont{Saha}},
  \bibinfo{author}{\bibfnamefont{A.}~\bibnamefont{Koda}}, \bibnamefont{et~al.},
  \bibinfo{journal}{Physica B: Condensed Matter}
  \textbf{\bibinfo{volume}{404}}, \bibinfo{pages}{749 } (\bibinfo{year}{2009}),
  ISSN \bibinfo{issn}{0921-4526},
  \urlprefix\url{http://www.sciencedirect.com/science/article/pii/S0921452608006613}.

\bibitem[{\citenamefont{Jeon et~al.}(2016)\citenamefont{Jeon, Huang, Yazici,
  Kanchanavatee, White, Ho, Jang, Pouse, and Maple}}]{MaplePRB16}
\bibinfo{author}{\bibfnamefont{I.}~\bibnamefont{Jeon}},
  \bibinfo{author}{\bibfnamefont{K.}~\bibnamefont{Huang}},
  \bibinfo{author}{\bibfnamefont{D.}~\bibnamefont{Yazici}},
  \bibinfo{author}{\bibfnamefont{N.}~\bibnamefont{Kanchanavatee}},
  \bibinfo{author}{\bibfnamefont{B.~D.} \bibnamefont{White}},
  \bibinfo{author}{\bibfnamefont{P.-C.} \bibnamefont{Ho}},
  \bibinfo{author}{\bibfnamefont{S.}~\bibnamefont{Jang}},
  \bibinfo{author}{\bibfnamefont{N.}~\bibnamefont{Pouse}}, \bibnamefont{and}
  \bibinfo{author}{\bibfnamefont{M.~B.} \bibnamefont{Maple}},
  \bibinfo{journal}{Phys. Rev. B} \textbf{\bibinfo{volume}{93}},
  \bibinfo{pages}{104507} (\bibinfo{year}{2016}),
  \urlprefix\url{https://link.aps.org/doi/10.1103/PhysRevB.93.104507}.

\bibitem[{\citenamefont{Yuhasz et~al.}(2006)\citenamefont{Yuhasz, Butch,
  Sayles, Ho, Jeffries, Yanagisawa, Frederick, Maple, Henkie, Pietraszko
  et~al.}}]{MAGStates_PhysRevB.73.144409}
\bibinfo{author}{\bibfnamefont{W.~M.} \bibnamefont{Yuhasz}},
  \bibinfo{author}{\bibfnamefont{N.~P.} \bibnamefont{Butch}},
  \bibinfo{author}{\bibfnamefont{T.~A.} \bibnamefont{Sayles}},
  \bibinfo{author}{\bibfnamefont{P.-C.} \bibnamefont{Ho}},
  \bibinfo{author}{\bibfnamefont{J.~R.} \bibnamefont{Jeffries}},
  \bibinfo{author}{\bibfnamefont{T.}~\bibnamefont{Yanagisawa}},
  \bibinfo{author}{\bibfnamefont{N.~A.} \bibnamefont{Frederick}},
  \bibinfo{author}{\bibfnamefont{M.~B.} \bibnamefont{Maple}},
  \bibinfo{author}{\bibfnamefont{Z.}~\bibnamefont{Henkie}},
  \bibinfo{author}{\bibfnamefont{A.}~\bibnamefont{Pietraszko}},
  \bibnamefont{et~al.}, \bibinfo{journal}{Phys. Rev. B}
  \textbf{\bibinfo{volume}{73}}, \bibinfo{pages}{144409}
  (\bibinfo{year}{2006}),
  \urlprefix\url{https://link.aps.org/doi/10.1103/PhysRevB.73.144409}.

\bibitem[{\citenamefont{Morelli and Meisner}(1995)}]{TherMat1_Morelli_1995}
\bibinfo{author}{\bibfnamefont{D.~T.} \bibnamefont{Morelli}} \bibnamefont{and}
  \bibinfo{author}{\bibfnamefont{G.~P.} \bibnamefont{Meisner}},
  \bibinfo{journal}{Journal of Applied Physics} \textbf{\bibinfo{volume}{77}},
  \bibinfo{pages}{3777} (\bibinfo{year}{1995}).

\bibitem[{\citenamefont{Sales et~al.}(1997)\citenamefont{Sales, Mandrus,
  Chakoumakos, Keppens, and Thompson}}]{TherMat2_Sales_1997_PhysRevB.56.15081}
\bibinfo{author}{\bibfnamefont{B.~C.} \bibnamefont{Sales}},
  \bibinfo{author}{\bibfnamefont{D.}~\bibnamefont{Mandrus}},
  \bibinfo{author}{\bibfnamefont{B.~C.} \bibnamefont{Chakoumakos}},
  \bibinfo{author}{\bibfnamefont{V.}~\bibnamefont{Keppens}}, \bibnamefont{and}
  \bibinfo{author}{\bibfnamefont{J.~R.} \bibnamefont{Thompson}},
  \bibinfo{journal}{Phys. Rev. B} \textbf{\bibinfo{volume}{56}},
  \bibinfo{pages}{15081} (\bibinfo{year}{1997}),
  \urlprefix\url{https://link.aps.org/doi/10.1103/PhysRevB.56.15081}.

\bibitem[{\citenamefont{Nolas et~al.}(1998)\citenamefont{Nolas, Cohn, and
  Slack}}]{TherMat3_Nolas_1998_PhysRevB.58.164}
\bibinfo{author}{\bibfnamefont{G.~S.} \bibnamefont{Nolas}},
  \bibinfo{author}{\bibfnamefont{J.~L.} \bibnamefont{Cohn}}, \bibnamefont{and}
  \bibinfo{author}{\bibfnamefont{G.~A.} \bibnamefont{Slack}},
  \bibinfo{journal}{Phys. Rev. B} \textbf{\bibinfo{volume}{58}},
  \bibinfo{pages}{164} (\bibinfo{year}{1998}),
  \urlprefix\url{https://link.aps.org/doi/10.1103/PhysRevB.58.164}.

\bibitem[{\citenamefont{Sales et~al.}(1996)\citenamefont{Sales, Mandrus, and
  Williams}}]{TherMat4_Sales1325_1996}
\bibinfo{author}{\bibfnamefont{B.~C.} \bibnamefont{Sales}},
  \bibinfo{author}{\bibfnamefont{D.}~\bibnamefont{Mandrus}}, \bibnamefont{and}
  \bibinfo{author}{\bibfnamefont{R.~K.} \bibnamefont{Williams}},
  \bibinfo{journal}{Science} \textbf{\bibinfo{volume}{272}},
  \bibinfo{pages}{1325} (\bibinfo{year}{1996}), ISSN \bibinfo{issn}{0036-8075},
  \eprint{http://science.sciencemag.org/content/272/5266/1325.full.pdf},
  \urlprefix\url{http://science.sciencemag.org/content/272/5266/1325}.

\bibitem[{\citenamefont{Gumeniuk et~al.}(2008)\citenamefont{Gumeniuk, Schnelle,
  Rosner, Nicklas, Leithe-Jasper, and Grin}}]{GumeniukPRL08}
\bibinfo{author}{\bibfnamefont{R.}~\bibnamefont{Gumeniuk}},
  \bibinfo{author}{\bibfnamefont{W.}~\bibnamefont{Schnelle}},
  \bibinfo{author}{\bibfnamefont{H.}~\bibnamefont{Rosner}},
  \bibinfo{author}{\bibfnamefont{M.}~\bibnamefont{Nicklas}},
  \bibinfo{author}{\bibfnamefont{A.}~\bibnamefont{Leithe-Jasper}},
  \bibnamefont{and} \bibinfo{author}{\bibfnamefont{Y.}~\bibnamefont{Grin}},
  \bibinfo{journal}{Phys. Rev. Lett.} \textbf{\bibinfo{volume}{100}},
  \bibinfo{pages}{017002} (\bibinfo{year}{2008}),
  \urlprefix\url{https://link.aps.org/doi/10.1103/PhysRevLett.100.017002}.

\bibitem[{\citenamefont{Gumeniuk et~al.}(2011)\citenamefont{Gumeniuk,
  Kvashnina, Schnelle, Nicklas, Borrmann, Rosner, Skourski, Tsirlin,
  Leithe-Jasper, and Grin}}]{GumeniukJP11}
\bibinfo{author}{\bibfnamefont{R.}~\bibnamefont{Gumeniuk}},
  \bibinfo{author}{\bibfnamefont{K.~O.} \bibnamefont{Kvashnina}},
  \bibinfo{author}{\bibfnamefont{W.}~\bibnamefont{Schnelle}},
  \bibinfo{author}{\bibfnamefont{M.}~\bibnamefont{Nicklas}},
  \bibinfo{author}{\bibfnamefont{H.}~\bibnamefont{Borrmann}},
  \bibinfo{author}{\bibfnamefont{H.}~\bibnamefont{Rosner}},
  \bibinfo{author}{\bibfnamefont{Y.}~\bibnamefont{Skourski}},
  \bibinfo{author}{\bibfnamefont{A.~A.} \bibnamefont{Tsirlin}},
  \bibinfo{author}{\bibfnamefont{A.}~\bibnamefont{Leithe-Jasper}},
  \bibnamefont{and} \bibinfo{author}{\bibfnamefont{Y.}~\bibnamefont{Grin}},
  \bibinfo{journal}{Journal of Physics: Condensed Matter}
  \textbf{\bibinfo{volume}{23}}, \bibinfo{pages}{465601}
  (\bibinfo{year}{2011}),
  \urlprefix\url{http://stacks.iop.org/0953-8984/23/i=46/a=465601}.

\bibitem[{\citenamefont{Chandra et~al.}(2016)\citenamefont{Chandra,
  Chattopadhyay, Roy, and Pandey}}]{ChandraPhilmag16}
\bibinfo{author}{\bibfnamefont{L.~S.~S.} \bibnamefont{Chandra}},
  \bibinfo{author}{\bibfnamefont{M.~K.} \bibnamefont{Chattopadhyay}},
  \bibinfo{author}{\bibfnamefont{S.~B.} \bibnamefont{Roy}}, \bibnamefont{and}
  \bibinfo{author}{\bibfnamefont{S.~K.} \bibnamefont{Pandey}},
  \bibinfo{journal}{Philosophical Magazine} \textbf{\bibinfo{volume}{96}},
  \bibinfo{pages}{2161} (\bibinfo{year}{2016}),
  \eprint{https://doi.org/10.1080/14786435.2016.1192722},
  \urlprefix\url{https://doi.org/10.1080/14786435.2016.1192722}.

\bibitem[{\citenamefont{Zhang et~al.}(2013)\citenamefont{Zhang, Chen, Jiao,
  Gumeniuk, Nicklas, Chen, Yang, Fu, Schnelle, Rosner
  et~al.}}]{Zhang-GumenPRB13}
\bibinfo{author}{\bibfnamefont{J.~L.} \bibnamefont{Zhang}},
  \bibinfo{author}{\bibfnamefont{Y.}~\bibnamefont{Chen}},
  \bibinfo{author}{\bibfnamefont{L.}~\bibnamefont{Jiao}},
  \bibinfo{author}{\bibfnamefont{R.}~\bibnamefont{Gumeniuk}},
  \bibinfo{author}{\bibfnamefont{M.}~\bibnamefont{Nicklas}},
  \bibinfo{author}{\bibfnamefont{Y.~H.} \bibnamefont{Chen}},
  \bibinfo{author}{\bibfnamefont{L.}~\bibnamefont{Yang}},
  \bibinfo{author}{\bibfnamefont{B.~H.} \bibnamefont{Fu}},
  \bibinfo{author}{\bibfnamefont{W.}~\bibnamefont{Schnelle}},
  \bibinfo{author}{\bibfnamefont{H.}~\bibnamefont{Rosner}},
  \bibnamefont{et~al.}, \bibinfo{journal}{Phys. Rev. B}
  \textbf{\bibinfo{volume}{87}}, \bibinfo{pages}{064502}
  (\bibinfo{year}{2013}),
  \urlprefix\url{https://link.aps.org/doi/10.1103/PhysRevB.87.064502}.

\bibitem[{\citenamefont{Peters et~al.}(2014)\citenamefont{Peters, Di~Marco,
  Thunstr\"om, Katsnelson, Kirilyuk, and
  Eriksson}}]{DFT_problems_PhysRevB.89.205109}
\bibinfo{author}{\bibfnamefont{L.}~\bibnamefont{Peters}},
  \bibinfo{author}{\bibfnamefont{I.}~\bibnamefont{Di~Marco}},
  \bibinfo{author}{\bibfnamefont{P.}~\bibnamefont{Thunstr\"om}},
  \bibinfo{author}{\bibfnamefont{M.~I.} \bibnamefont{Katsnelson}},
  \bibinfo{author}{\bibfnamefont{A.}~\bibnamefont{Kirilyuk}}, \bibnamefont{and}
  \bibinfo{author}{\bibfnamefont{O.}~\bibnamefont{Eriksson}},
  \bibinfo{journal}{Phys. Rev. B} \textbf{\bibinfo{volume}{89}},
  \bibinfo{pages}{205109} (\bibinfo{year}{2014}),
  \urlprefix\url{https://link.aps.org/doi/10.1103/PhysRevB.89.205109}.

\bibitem[{\citenamefont{Petit et~al.}(2016)\citenamefont{Petit, Szotek,
  Lüders, and Svane}}]{DFT_problems_Petit_2016}
\bibinfo{author}{\bibfnamefont{L.}~\bibnamefont{Petit}},
  \bibinfo{author}{\bibfnamefont{Z.}~\bibnamefont{Szotek}},
  \bibinfo{author}{\bibfnamefont{M.}~\bibnamefont{Lüders}}, \bibnamefont{and}
  \bibinfo{author}{\bibfnamefont{A.}~\bibnamefont{Svane}},
  \bibinfo{journal}{Journal of Physics: Condensed Matter}
  \textbf{\bibinfo{volume}{28}}, \bibinfo{pages}{223001}
  (\bibinfo{year}{2016}),
  \urlprefix\url{https://doi.org/10.1088%2F0953-8984%2F28%2F22%2F223001}.

\bibitem[{\citenamefont{Xi}(2008)}]{MgB2_Xi_2008}
\bibinfo{author}{\bibfnamefont{X.~X.} \bibnamefont{Xi}},
  \bibinfo{journal}{Reports on Progress in Physics}
  \textbf{\bibinfo{volume}{71}}, \bibinfo{pages}{116501}
  (\bibinfo{year}{2008}),
  \urlprefix\url{https://doi.org/10.1088%2F0034-4885%2F71%2F11%2F116501}.

\bibitem[{\citenamefont{Kortus et~al.}(2001)\citenamefont{Kortus, Mazin,
  Belashchenko, Antropov, and Boyer}}]{MgB2_PhysRevLett.86.4656}
\bibinfo{author}{\bibfnamefont{J.}~\bibnamefont{Kortus}},
  \bibinfo{author}{\bibfnamefont{I.~I.} \bibnamefont{Mazin}},
  \bibinfo{author}{\bibfnamefont{K.~D.} \bibnamefont{Belashchenko}},
  \bibinfo{author}{\bibfnamefont{V.~P.} \bibnamefont{Antropov}},
  \bibnamefont{and} \bibinfo{author}{\bibfnamefont{L.~L.} \bibnamefont{Boyer}},
  \bibinfo{journal}{Phys. Rev. Lett.} \textbf{\bibinfo{volume}{86}},
  \bibinfo{pages}{4656} (\bibinfo{year}{2001}),
  \urlprefix\url{https://link.aps.org/doi/10.1103/PhysRevLett.86.4656}.

\bibitem[{\citenamefont{Floris et~al.}(2007)\citenamefont{Floris, Sanna,
  Massidda, and Gross}}]{Pd_PhysRevB.75.054508}
\bibinfo{author}{\bibfnamefont{A.}~\bibnamefont{Floris}},
  \bibinfo{author}{\bibfnamefont{A.}~\bibnamefont{Sanna}},
  \bibinfo{author}{\bibfnamefont{S.}~\bibnamefont{Massidda}}, \bibnamefont{and}
  \bibinfo{author}{\bibfnamefont{E.~K.~U.} \bibnamefont{Gross}},
  \bibinfo{journal}{Phys. Rev. B} \textbf{\bibinfo{volume}{75}},
  \bibinfo{pages}{054508} (\bibinfo{year}{2007}),
  \urlprefix\url{https://link.aps.org/doi/10.1103/PhysRevB.75.054508}.

\bibitem[{\citenamefont{Blackford and March}(1969)}]{Pd_PhysRev.186.397}
\bibinfo{author}{\bibfnamefont{B.~L.} \bibnamefont{Blackford}}
  \bibnamefont{and} \bibinfo{author}{\bibfnamefont{R.~H.} \bibnamefont{March}},
  \bibinfo{journal}{Phys. Rev.} \textbf{\bibinfo{volume}{186}},
  \bibinfo{pages}{397} (\bibinfo{year}{1969}),
  \urlprefix\url{https://link.aps.org/doi/10.1103/PhysRev.186.397}.

\bibitem[{\citenamefont{Bauer et~al.}(2007)\citenamefont{Bauer, Grytsiv, Chen,
  Melnychenko-Koblyuk, Hilscher, Kaldarar, Michor, Royanian, Giester, Rotter
  et~al.}}]{BauerPRL07}
\bibinfo{author}{\bibfnamefont{E.}~\bibnamefont{Bauer}},
  \bibinfo{author}{\bibfnamefont{A.}~\bibnamefont{Grytsiv}},
  \bibinfo{author}{\bibfnamefont{X.-Q.} \bibnamefont{Chen}},
  \bibinfo{author}{\bibfnamefont{N.}~\bibnamefont{Melnychenko-Koblyuk}},
  \bibinfo{author}{\bibfnamefont{G.}~\bibnamefont{Hilscher}},
  \bibinfo{author}{\bibfnamefont{H.}~\bibnamefont{Kaldarar}},
  \bibinfo{author}{\bibfnamefont{H.}~\bibnamefont{Michor}},
  \bibinfo{author}{\bibfnamefont{E.}~\bibnamefont{Royanian}},
  \bibinfo{author}{\bibfnamefont{G.}~\bibnamefont{Giester}},
  \bibinfo{author}{\bibfnamefont{M.}~\bibnamefont{Rotter}},
  \bibnamefont{et~al.}, \bibinfo{journal}{Phys. Rev. Lett.}
  \textbf{\bibinfo{volume}{99}}, \bibinfo{pages}{217001}
  (\bibinfo{year}{2007}),
  \urlprefix\url{https://link.aps.org/doi/10.1103/PhysRevLett.99.217001}.

\bibitem[{\citenamefont{Rosner et~al.}(2009)\citenamefont{Rosner, Gegner,
  Regesch, Schnelle, Gumeniuk, Leithe-Jasper, Fujiwara, Haupricht, Koethe,
  Hsieh et~al.}}]{RosnerPRB09}
\bibinfo{author}{\bibfnamefont{H.}~\bibnamefont{Rosner}},
  \bibinfo{author}{\bibfnamefont{J.}~\bibnamefont{Gegner}},
  \bibinfo{author}{\bibfnamefont{D.}~\bibnamefont{Regesch}},
  \bibinfo{author}{\bibfnamefont{W.}~\bibnamefont{Schnelle}},
  \bibinfo{author}{\bibfnamefont{R.}~\bibnamefont{Gumeniuk}},
  \bibinfo{author}{\bibfnamefont{A.}~\bibnamefont{Leithe-Jasper}},
  \bibinfo{author}{\bibfnamefont{H.}~\bibnamefont{Fujiwara}},
  \bibinfo{author}{\bibfnamefont{T.}~\bibnamefont{Haupricht}},
  \bibinfo{author}{\bibfnamefont{T.~C.} \bibnamefont{Koethe}},
  \bibinfo{author}{\bibfnamefont{H.-H.} \bibnamefont{Hsieh}},
  \bibnamefont{et~al.}, \bibinfo{journal}{Phys. Rev. B}
  \textbf{\bibinfo{volume}{80}}, \bibinfo{pages}{075114}
  (\bibinfo{year}{2009}),
  \urlprefix\url{https://link.aps.org/doi/10.1103/PhysRevB.80.075114}.

\bibitem[{\citenamefont{Gumeniuk
  et~al.}(2010{\natexlab{b}})\citenamefont{Gumeniuk, SchÃ¶neich,
  Leithe-Jasper, Schnelle, Nicklas, Rosner, Ormeci, Burkhardt, Schmidt, Schwarz
  et~al.}}]{GumeniukNewjPhys10}
\bibinfo{author}{\bibfnamefont{R.}~\bibnamefont{Gumeniuk}},
  \bibinfo{author}{\bibfnamefont{M.}~\bibnamefont{SchÃ¶neich}},
  \bibinfo{author}{\bibfnamefont{A.}~\bibnamefont{Leithe-Jasper}},
  \bibinfo{author}{\bibfnamefont{W.}~\bibnamefont{Schnelle}},
  \bibinfo{author}{\bibfnamefont{M.}~\bibnamefont{Nicklas}},
  \bibinfo{author}{\bibfnamefont{H.}~\bibnamefont{Rosner}},
  \bibinfo{author}{\bibfnamefont{A.}~\bibnamefont{Ormeci}},
  \bibinfo{author}{\bibfnamefont{U.}~\bibnamefont{Burkhardt}},
  \bibinfo{author}{\bibfnamefont{M.}~\bibnamefont{Schmidt}},
  \bibinfo{author}{\bibfnamefont{U.}~\bibnamefont{Schwarz}},
  \bibnamefont{et~al.}, \bibinfo{journal}{New Journal of Physics}
  \textbf{\bibinfo{volume}{12}}, \bibinfo{pages}{103035}
  (\bibinfo{year}{2010}{\natexlab{b}}),
  \urlprefix\url{http://stacks.iop.org/1367-2630/12/i=10/a=103035}.

\bibitem[{\citenamefont{Pfau et~al.}(2016)\citenamefont{Pfau, Nicklas,
  Stockert, Gumeniuk, Schnelle, Leithe-Jasper, Grin, and
  Steglich}}]{SteglichPRB16}
\bibinfo{author}{\bibfnamefont{H.}~\bibnamefont{Pfau}},
  \bibinfo{author}{\bibfnamefont{M.}~\bibnamefont{Nicklas}},
  \bibinfo{author}{\bibfnamefont{U.}~\bibnamefont{Stockert}},
  \bibinfo{author}{\bibfnamefont{R.}~\bibnamefont{Gumeniuk}},
  \bibinfo{author}{\bibfnamefont{W.}~\bibnamefont{Schnelle}},
  \bibinfo{author}{\bibfnamefont{A.}~\bibnamefont{Leithe-Jasper}},
  \bibinfo{author}{\bibfnamefont{Y.}~\bibnamefont{Grin}}, \bibnamefont{and}
  \bibinfo{author}{\bibfnamefont{F.}~\bibnamefont{Steglich}},
  \bibinfo{journal}{Phys. Rev. B} \textbf{\bibinfo{volume}{94}},
  \bibinfo{pages}{054523} (\bibinfo{year}{2016}),
  \urlprefix\url{https://link.aps.org/doi/10.1103/PhysRevB.94.054523}.

\bibitem[{\citenamefont{Chandra et~al.}(2012)\citenamefont{Chandra,
  Chattopadhyay, and Roy}}]{ChandraPhilmag12}
\bibinfo{author}{\bibfnamefont{L.~S.} \bibnamefont{Chandra}},
  \bibinfo{author}{\bibfnamefont{M.}~\bibnamefont{Chattopadhyay}},
  \bibnamefont{and} \bibinfo{author}{\bibfnamefont{S.}~\bibnamefont{Roy}},
  \bibinfo{journal}{Philosophical Magazine} \textbf{\bibinfo{volume}{92}},
  \bibinfo{pages}{3866} (\bibinfo{year}{2012}),
  \eprint{https://doi.org/10.1080/14786435.2012.691218},
  \urlprefix\url{https://doi.org/10.1080/14786435.2012.691218}.

\bibitem[{\citenamefont{Nakamura
  et~al.}(2010{\natexlab{a}})\citenamefont{Nakamura, Okazaki, Yoshida, Wakita,
  Hirai, Muraoka, Takeya, Hirata, Kumigashira, Oshima et~al.}}]{NakamuraJPSJ10}
\bibinfo{author}{\bibfnamefont{Y.}~\bibnamefont{Nakamura}},
  \bibinfo{author}{\bibfnamefont{H.}~\bibnamefont{Okazaki}},
  \bibinfo{author}{\bibfnamefont{R.}~\bibnamefont{Yoshida}},
  \bibinfo{author}{\bibfnamefont{T.}~\bibnamefont{Wakita}},
  \bibinfo{author}{\bibfnamefont{M.}~\bibnamefont{Hirai}},
  \bibinfo{author}{\bibfnamefont{Y.}~\bibnamefont{Muraoka}},
  \bibinfo{author}{\bibfnamefont{H.}~\bibnamefont{Takeya}},
  \bibinfo{author}{\bibfnamefont{K.}~\bibnamefont{Hirata}},
  \bibinfo{author}{\bibfnamefont{H.}~\bibnamefont{Kumigashira}},
  \bibinfo{author}{\bibfnamefont{M.}~\bibnamefont{Oshima}},
  \bibnamefont{et~al.}, \bibinfo{journal}{Journal of the Physical Society of
  Japan} \textbf{\bibinfo{volume}{79}}, \bibinfo{pages}{124701}
  (\bibinfo{year}{2010}{\natexlab{a}}),
  \eprint{https://doi.org/10.1143/JPSJ.79.124701},
  \urlprefix\url{https://doi.org/10.1143/JPSJ.79.124701}.

\bibitem[{\citenamefont{Nakamura et~al.}(2012)\citenamefont{Nakamura, Okazaki,
  Yoshida, Wakita, Takeya, Hirata, Hirai, Muraoka, and Yokoya}}]{NakamuraPRB12}
\bibinfo{author}{\bibfnamefont{Y.}~\bibnamefont{Nakamura}},
  \bibinfo{author}{\bibfnamefont{H.}~\bibnamefont{Okazaki}},
  \bibinfo{author}{\bibfnamefont{R.}~\bibnamefont{Yoshida}},
  \bibinfo{author}{\bibfnamefont{T.}~\bibnamefont{Wakita}},
  \bibinfo{author}{\bibfnamefont{H.}~\bibnamefont{Takeya}},
  \bibinfo{author}{\bibfnamefont{K.}~\bibnamefont{Hirata}},
  \bibinfo{author}{\bibfnamefont{M.}~\bibnamefont{Hirai}},
  \bibinfo{author}{\bibfnamefont{Y.}~\bibnamefont{Muraoka}}, \bibnamefont{and}
  \bibinfo{author}{\bibfnamefont{T.}~\bibnamefont{Yokoya}},
  \bibinfo{journal}{Phys. Rev. B} \textbf{\bibinfo{volume}{86}},
  \bibinfo{pages}{014521} (\bibinfo{year}{2012}),
  \urlprefix\url{https://link.aps.org/doi/10.1103/PhysRevB.86.014521}.

\bibitem[{\citenamefont{Gurevich}(2007)}]{GurevichPhysica07}
\bibinfo{author}{\bibfnamefont{A.}~\bibnamefont{Gurevich}},
  \bibinfo{journal}{Physica C: Superconductivity}
  \textbf{\bibinfo{volume}{456}}, \bibinfo{pages}{160 } (\bibinfo{year}{2007}),
  ISSN \bibinfo{issn}{0921-4534}, \bibinfo{note}{recent Advances in MgB2
  Research},
  \urlprefix\url{http://www.sciencedirect.com/science/article/pii/S0921453407000111}.

\bibitem[{\citenamefont{Tsuda et~al.}(2004)\citenamefont{Tsuda, Yokoya, Shin,
  Takano, Kito, Matsushita, Yin, Itoh, and Harima}}]{MgB2_TSUDA200436}
\bibinfo{author}{\bibfnamefont{S.}~\bibnamefont{Tsuda}},
  \bibinfo{author}{\bibfnamefont{T.}~\bibnamefont{Yokoya}},
  \bibinfo{author}{\bibfnamefont{S.}~\bibnamefont{Shin}},
  \bibinfo{author}{\bibfnamefont{Y.}~\bibnamefont{Takano}},
  \bibinfo{author}{\bibfnamefont{H.}~\bibnamefont{Kito}},
  \bibinfo{author}{\bibfnamefont{A.}~\bibnamefont{Matsushita}},
  \bibinfo{author}{\bibfnamefont{F.}~\bibnamefont{Yin}},
  \bibinfo{author}{\bibfnamefont{J.}~\bibnamefont{Itoh}}, \bibnamefont{and}
  \bibinfo{author}{\bibfnamefont{H.}~\bibnamefont{Harima}},
  \bibinfo{journal}{Physica C: Superconductivity}
  \textbf{\bibinfo{volume}{412-414}}, \bibinfo{pages}{36 }
  (\bibinfo{year}{2004}), ISSN \bibinfo{issn}{0921-4534},
  \bibinfo{note}{proceedings of the 16th International Symposium on
  Superconductivity (ISS 2003). Advances in Superconductivity XVI. Part I},
  \urlprefix\url{http://www.sciencedirect.com/science/article/pii/S0921453404006173}.

\bibitem[{\citenamefont{Tsuda et~al.}(2007)\citenamefont{Tsuda, Yokoya, and
  Shin}}]{MgB2_TSUDA2007126}
\bibinfo{author}{\bibfnamefont{S.}~\bibnamefont{Tsuda}},
  \bibinfo{author}{\bibfnamefont{T.}~\bibnamefont{Yokoya}}, \bibnamefont{and}
  \bibinfo{author}{\bibfnamefont{S.}~\bibnamefont{Shin}},
  \bibinfo{journal}{Physica C: Superconductivity}
  \textbf{\bibinfo{volume}{456}}, \bibinfo{pages}{126 } (\bibinfo{year}{2007}),
  ISSN \bibinfo{issn}{0921-4534}, \bibinfo{note}{recent Advances in MgB2
  Research},
  \urlprefix\url{http://www.sciencedirect.com/science/article/pii/S0921453406008598}.

\bibitem[{\citenamefont{Lykken et~al.}(1971)\citenamefont{Lykken, Geiger, Dy,
  and Mitchell}}]{Pd_PhysRevB.4.1523}
\bibinfo{author}{\bibfnamefont{G.~I.} \bibnamefont{Lykken}},
  \bibinfo{author}{\bibfnamefont{A.~L.} \bibnamefont{Geiger}},
  \bibinfo{author}{\bibfnamefont{K.~S.} \bibnamefont{Dy}}, \bibnamefont{and}
  \bibinfo{author}{\bibfnamefont{E.~N.} \bibnamefont{Mitchell}},
  \bibinfo{journal}{Phys. Rev. B} \textbf{\bibinfo{volume}{4}},
  \bibinfo{pages}{1523} (\bibinfo{year}{1971}),
  \urlprefix\url{https://link.aps.org/doi/10.1103/PhysRevB.4.1523}.

\bibitem[{\citenamefont{Singh et~al.}(2016)\citenamefont{Singh, Adhikari,
  Zhang, Huang, Yazici, Jeon, Maple, Dzero, and Almasan}}]{SinghPRB16}
\bibinfo{author}{\bibfnamefont{Y.~P.} \bibnamefont{Singh}},
  \bibinfo{author}{\bibfnamefont{R.~B.} \bibnamefont{Adhikari}},
  \bibinfo{author}{\bibfnamefont{S.}~\bibnamefont{Zhang}},
  \bibinfo{author}{\bibfnamefont{K.}~\bibnamefont{Huang}},
  \bibinfo{author}{\bibfnamefont{D.}~\bibnamefont{Yazici}},
  \bibinfo{author}{\bibfnamefont{I.}~\bibnamefont{Jeon}},
  \bibinfo{author}{\bibfnamefont{M.~B.} \bibnamefont{Maple}},
  \bibinfo{author}{\bibfnamefont{M.}~\bibnamefont{Dzero}}, \bibnamefont{and}
  \bibinfo{author}{\bibfnamefont{C.~C.} \bibnamefont{Almasan}},
  \bibinfo{journal}{Phys. Rev. B} \textbf{\bibinfo{volume}{94}},
  \bibinfo{pages}{144502} (\bibinfo{year}{2016}),
  \urlprefix\url{https://link.aps.org/doi/10.1103/PhysRevB.94.144502}.

\bibitem[{\citenamefont{Huang et~al.}(2014)\citenamefont{Huang, Shu, Lum,
  White, Janoschek, Yazici, Hamlin, Zocco, Ho, Baumbach et~al.}}]{MaplePRB14}
\bibinfo{author}{\bibfnamefont{K.}~\bibnamefont{Huang}},
  \bibinfo{author}{\bibfnamefont{L.}~\bibnamefont{Shu}},
  \bibinfo{author}{\bibfnamefont{I.~K.} \bibnamefont{Lum}},
  \bibinfo{author}{\bibfnamefont{B.~D.} \bibnamefont{White}},
  \bibinfo{author}{\bibfnamefont{M.}~\bibnamefont{Janoschek}},
  \bibinfo{author}{\bibfnamefont{D.}~\bibnamefont{Yazici}},
  \bibinfo{author}{\bibfnamefont{J.~J.} \bibnamefont{Hamlin}},
  \bibinfo{author}{\bibfnamefont{D.~A.} \bibnamefont{Zocco}},
  \bibinfo{author}{\bibfnamefont{P.-C.} \bibnamefont{Ho}},
  \bibinfo{author}{\bibfnamefont{R.~E.} \bibnamefont{Baumbach}},
  \bibnamefont{et~al.}, \bibinfo{journal}{Phys. Rev. B}
  \textbf{\bibinfo{volume}{89}}, \bibinfo{pages}{035145}
  (\bibinfo{year}{2014}),
  \urlprefix\url{https://link.aps.org/doi/10.1103/PhysRevB.89.035145}.

\bibitem[{\citenamefont{Kanetake et~al.}(2010)\citenamefont{Kanetake, Mukuda,
  Kitaoka, ichi Magishi, Sugawara, Itoh, and Haller}}]{KanatakeJPS10}
\bibinfo{author}{\bibfnamefont{F.}~\bibnamefont{Kanetake}},
  \bibinfo{author}{\bibfnamefont{H.}~\bibnamefont{Mukuda}},
  \bibinfo{author}{\bibfnamefont{Y.}~\bibnamefont{Kitaoka}},
  \bibinfo{author}{\bibfnamefont{K.}~\bibnamefont{ichi Magishi}},
  \bibinfo{author}{\bibfnamefont{H.}~\bibnamefont{Sugawara}},
  \bibinfo{author}{\bibfnamefont{K.~M.} \bibnamefont{Itoh}}, \bibnamefont{and}
  \bibinfo{author}{\bibfnamefont{E.~E.} \bibnamefont{Haller}},
  \bibinfo{journal}{Journal of the Physical Society of Japan}
  \textbf{\bibinfo{volume}{79}}, \bibinfo{pages}{063702}
  (\bibinfo{year}{2010}), \eprint{https://doi.org/10.1143/JPSJ.79.063702},
  \urlprefix\url{https://doi.org/10.1143/JPSJ.79.063702}.

\bibitem[{\citenamefont{Nicklas et~al.}(2012)\citenamefont{Nicklas, Kirchner,
  Borth, Gumeniuk, Schnelle, Rosner, Borrmann, Leithe-Jasper, Grin, and
  Steglich}}]{Ce_Mag_PhysRevLett.109.236405}
\bibinfo{author}{\bibfnamefont{M.}~\bibnamefont{Nicklas}},
  \bibinfo{author}{\bibfnamefont{S.}~\bibnamefont{Kirchner}},
  \bibinfo{author}{\bibfnamefont{R.}~\bibnamefont{Borth}},
  \bibinfo{author}{\bibfnamefont{R.}~\bibnamefont{Gumeniuk}},
  \bibinfo{author}{\bibfnamefont{W.}~\bibnamefont{Schnelle}},
  \bibinfo{author}{\bibfnamefont{H.}~\bibnamefont{Rosner}},
  \bibinfo{author}{\bibfnamefont{H.}~\bibnamefont{Borrmann}},
  \bibinfo{author}{\bibfnamefont{A.}~\bibnamefont{Leithe-Jasper}},
  \bibinfo{author}{\bibfnamefont{Y.}~\bibnamefont{Grin}}, \bibnamefont{and}
  \bibinfo{author}{\bibfnamefont{F.}~\bibnamefont{Steglich}},
  \bibinfo{journal}{Phys. Rev. Lett.} \textbf{\bibinfo{volume}{109}},
  \bibinfo{pages}{236405} (\bibinfo{year}{2012}),
  \urlprefix\url{https://link.aps.org/doi/10.1103/PhysRevLett.109.236405}.

\bibitem[{\citenamefont{Humer et~al.}(2013)\citenamefont{Humer, Royanian,
  Michor, Bauer, Grytsiv, Chen, Podloucky, and
  Rogl}}]{thermoelectric_capabilities}
\bibinfo{author}{\bibfnamefont{S.}~\bibnamefont{Humer}},
  \bibinfo{author}{\bibfnamefont{E.}~\bibnamefont{Royanian}},
  \bibinfo{author}{\bibfnamefont{H.}~\bibnamefont{Michor}},
  \bibinfo{author}{\bibfnamefont{E.}~\bibnamefont{Bauer}},
  \bibinfo{author}{\bibfnamefont{A.}~\bibnamefont{Grytsiv}},
  \bibinfo{author}{\bibfnamefont{M.~X.} \bibnamefont{Chen}},
  \bibinfo{author}{\bibfnamefont{R.}~\bibnamefont{Podloucky}},
  \bibnamefont{and} \bibinfo{author}{\bibfnamefont{P.}~\bibnamefont{Rogl}},
  \emph{\bibinfo{title}{From Superconductivity Towards Thermoelectricity:
  Ge-Based Skutterudites. In: Zlatic V., Hewson A. (eds) New Materials for
  Thermoelectric Applications: Theory and Experiment. NATO Science for Peace
  and Security Series B: Physics and Biophysics. Springer, Dordrecht}}
  (\bibinfo{year}{2013}).

\bibitem[{\citenamefont{Slack}(1995)}]{PGEC1}
\bibinfo{author}{\bibfnamefont{G.~A.} \bibnamefont{Slack}},
  \emph{\bibinfo{title}{CRC Handbook of Thermoelectrics}}
  (\bibinfo{publisher}{Rowe, D. M.}, \bibinfo{year}{1995}).

\bibitem[{\citenamefont{Beekman et~al.}(2015)\citenamefont{Beekman, Morelli,
  and Nolas}}]{PGEC2}
\bibinfo{author}{\bibfnamefont{M.}~\bibnamefont{Beekman}},
  \bibinfo{author}{\bibfnamefont{D.~T.} \bibnamefont{Morelli}},
  \bibnamefont{and} \bibinfo{author}{\bibfnamefont{G.~S.} \bibnamefont{Nolas}},
  \bibinfo{journal}{Nature Materials} \textbf{\bibinfo{volume}{14}},
  \bibinfo{pages}{1182} (\bibinfo{year}{2015}).

\bibitem[{\citenamefont{Slack and Tsoukala}(1994)}]{PGEC3}
\bibinfo{author}{\bibfnamefont{G.~A.} \bibnamefont{Slack}} \bibnamefont{and}
  \bibinfo{author}{\bibfnamefont{V.~G.} \bibnamefont{Tsoukala}},
  \bibinfo{journal}{Journal of Applied Physics} \textbf{\bibinfo{volume}{76}},
  \bibinfo{pages}{1665} (\bibinfo{year}{1994}).

\bibitem[{\citenamefont{Nakamura
  et~al.}(2010{\natexlab{b}})\citenamefont{Nakamura, Okazaki, Yoshida, Wakita,
  Hirai, Muraoka, Takeya, Hirata, Kumigashira, Oshima et~al.}}]{NakamuraJPS10}
\bibinfo{author}{\bibfnamefont{Y.}~\bibnamefont{Nakamura}},
  \bibinfo{author}{\bibfnamefont{H.}~\bibnamefont{Okazaki}},
  \bibinfo{author}{\bibfnamefont{R.}~\bibnamefont{Yoshida}},
  \bibinfo{author}{\bibfnamefont{T.}~\bibnamefont{Wakita}},
  \bibinfo{author}{\bibfnamefont{M.}~\bibnamefont{Hirai}},
  \bibinfo{author}{\bibfnamefont{Y.}~\bibnamefont{Muraoka}},
  \bibinfo{author}{\bibfnamefont{H.}~\bibnamefont{Takeya}},
  \bibinfo{author}{\bibfnamefont{K.}~\bibnamefont{Hirata}},
  \bibinfo{author}{\bibfnamefont{H.}~\bibnamefont{Kumigashira}},
  \bibinfo{author}{\bibfnamefont{M.}~\bibnamefont{Oshima}},
  \bibnamefont{et~al.}, \bibinfo{journal}{J. Phys. Soc. Japan}
  \textbf{\bibinfo{volume}{79}}, \bibinfo{pages}{124701}
  (\bibinfo{year}{2010}{\natexlab{b}}),
  \urlprefix\url{http://dx.doi.org/10.1143/JPSJ.79.124701}.

\bibitem[{\citenamefont{Hohenberg and Kohn}(1964)}]{Hohn-Kohn}
\bibinfo{author}{\bibfnamefont{P.}~\bibnamefont{Hohenberg}} \bibnamefont{and}
  \bibinfo{author}{\bibfnamefont{W.}~\bibnamefont{Kohn}},
  \bibinfo{journal}{Phys. Rev. B} \textbf{\bibinfo{volume}{136}},
  \bibinfo{pages}{864} (\bibinfo{year}{1964}).

\bibitem[{\citenamefont{Kohn and Sham}(1965)}]{Kohn-Sham}
\bibinfo{author}{\bibfnamefont{W.}~\bibnamefont{Kohn}} \bibnamefont{and}
  \bibinfo{author}{\bibfnamefont{L.~J.} \bibnamefont{Sham}},
  \bibinfo{journal}{Phys. Rev. B} \textbf{\bibinfo{volume}{140}},
  \bibinfo{pages}{1133} (\bibinfo{year}{1965}).

\bibitem[{\citenamefont{Perdew et~al.}(1996)\citenamefont{Perdew, Burke, and
  Ernzerhof}}]{PBE}
\bibinfo{author}{\bibfnamefont{J.~P.} \bibnamefont{Perdew}},
  \bibinfo{author}{\bibfnamefont{K.}~\bibnamefont{Burke}}, \bibnamefont{and}
  \bibinfo{author}{\bibfnamefont{M.}~\bibnamefont{Ernzerhof}},
  \bibinfo{journal}{Phys. Rev. Lett.} \textbf{\bibinfo{volume}{77}},
  \bibinfo{pages}{3865} (\bibinfo{year}{1996}).

\bibitem[{\citenamefont{Kresse and Furthmuller}(1996)}]{Kresse96}
\bibinfo{author}{\bibfnamefont{G.}~\bibnamefont{Kresse}} \bibnamefont{and}
  \bibinfo{author}{\bibfnamefont{J.}~\bibnamefont{Furthmuller}},
  \bibinfo{journal}{Phys. Rev. B} \textbf{\bibinfo{volume}{54}},
  \bibinfo{pages}{11169} (\bibinfo{year}{1996}).

\bibitem[{\citenamefont{Kresse and Joubert}(1999)}]{Kresse99}
\bibinfo{author}{\bibfnamefont{G.}~\bibnamefont{Kresse}} \bibnamefont{and}
  \bibinfo{author}{\bibfnamefont{D.}~\bibnamefont{Joubert}},
  \bibinfo{journal}{Phys. Rev. B} \textbf{\bibinfo{volume}{59}},
  \bibinfo{pages}{1758} (\bibinfo{year}{1999}).

\bibitem[{\citenamefont{Blaha et~al.}(2018)\citenamefont{Blaha, Schwarz,
  Madsen, Kvasnicka, Luitz, Laskowski, Tran, and Marks}}]{wien2k}
\bibinfo{author}{\bibfnamefont{P.}~\bibnamefont{Blaha}},
  \bibinfo{author}{\bibfnamefont{K.}~\bibnamefont{Schwarz}},
  \bibinfo{author}{\bibfnamefont{G.~K.~H.} \bibnamefont{Madsen}},
  \bibinfo{author}{\bibfnamefont{D.}~\bibnamefont{Kvasnicka}},
  \bibinfo{author}{\bibfnamefont{J.}~\bibnamefont{Luitz}},
  \bibinfo{author}{\bibfnamefont{R.}~\bibnamefont{Laskowski}},
  \bibinfo{author}{\bibfnamefont{F.}~\bibnamefont{Tran}}, \bibnamefont{and}
  \bibinfo{author}{\bibfnamefont{L.~D.} \bibnamefont{Marks}},
  \emph{\bibinfo{title}{“WIEN2k, An Augmented Plane Wave + Local Orbitals
  Program for Calculating Crystal Properties” (Karlheinz Schwarz, Techn.
  Universit{\"a}t Wien, Austria). ISBN 3-9501031-1-2}} (\bibinfo{year}{2018}).

\bibitem[{\citenamefont{Setyawan and Curtarolo}(2010)}]{SETYAWAN2010299}
\bibinfo{author}{\bibfnamefont{W.}~\bibnamefont{Setyawan}} \bibnamefont{and}
  \bibinfo{author}{\bibfnamefont{S.}~\bibnamefont{Curtarolo}},
  \bibinfo{journal}{Computational Materials Science}
  \textbf{\bibinfo{volume}{49}}, \bibinfo{pages}{299 } (\bibinfo{year}{2010}),
  ISSN \bibinfo{issn}{0927-0256},
  \urlprefix\url{http://www.sciencedirect.com/science/article/pii/S0927025610002697}.

\bibitem[{\citenamefont{Schwarz et~al.}(0002)\citenamefont{Schwarz, Blaha, and
  Trickey}}]{FatBands}
\bibinfo{author}{\bibfnamefont{K.}~\bibnamefont{Schwarz}},
  \bibinfo{author}{\bibfnamefont{P.}~\bibnamefont{Blaha}}, \bibnamefont{and}
  \bibinfo{author}{\bibfnamefont{S.}~\bibnamefont{Trickey}},
  \bibinfo{journal}{Molecular Physics} \textbf{\bibinfo{volume}{108}},
  \bibinfo{pages}{21} (\bibinfo{year}{0002}).

\bibitem[{\citenamefont{Dudarev et~al.}(1998)\citenamefont{Dudarev, Botton,
  Savrasov, Humphreys, and Sutton}}]{U_PhysRevB.57.1505}
\bibinfo{author}{\bibfnamefont{S.~L.} \bibnamefont{Dudarev}},
  \bibinfo{author}{\bibfnamefont{G.~A.} \bibnamefont{Botton}},
  \bibinfo{author}{\bibfnamefont{S.~Y.} \bibnamefont{Savrasov}},
  \bibinfo{author}{\bibfnamefont{C.~J.} \bibnamefont{Humphreys}},
  \bibnamefont{and} \bibinfo{author}{\bibfnamefont{A.~P.}
  \bibnamefont{Sutton}}, \bibinfo{journal}{Phys. Rev. B}
  \textbf{\bibinfo{volume}{57}}, \bibinfo{pages}{1505} (\bibinfo{year}{1998}),
  \urlprefix\url{https://link.aps.org/doi/10.1103/PhysRevB.57.1505}.

\bibitem[{\citenamefont{Anisimov et~al.}(1993)\citenamefont{Anisimov, Solovyev,
  Korotin, Czy\ifmmode~\dot{z}\else \.{z}\fi{}yk, and
  Sawatzky}}]{U_PhysRevB.48.16929}
\bibinfo{author}{\bibfnamefont{V.~I.} \bibnamefont{Anisimov}},
  \bibinfo{author}{\bibfnamefont{I.~V.} \bibnamefont{Solovyev}},
  \bibinfo{author}{\bibfnamefont{M.~A.} \bibnamefont{Korotin}},
  \bibinfo{author}{\bibfnamefont{M.~T.} \bibnamefont{Czy\ifmmode~\dot{z}\else
  \.{z}\fi{}yk}}, \bibnamefont{and} \bibinfo{author}{\bibfnamefont{G.~A.}
  \bibnamefont{Sawatzky}}, \bibinfo{journal}{Phys. Rev. B}
  \textbf{\bibinfo{volume}{48}}, \bibinfo{pages}{16929} (\bibinfo{year}{1993}),
  \urlprefix\url{https://link.aps.org/doi/10.1103/PhysRevB.48.16929}.

\bibitem[{\citenamefont{Haule et~al.}(2010)\citenamefont{Haule, Yee, and
  Kim}}]{Haule_prb10}
\bibinfo{author}{\bibfnamefont{K.}~\bibnamefont{Haule}},
  \bibinfo{author}{\bibfnamefont{C.-H.} \bibnamefont{Yee}}, \bibnamefont{and}
  \bibinfo{author}{\bibfnamefont{K.}~\bibnamefont{Kim}},
  \bibinfo{journal}{Phys. Rev. B} \textbf{\bibinfo{volume}{81}},
  \bibinfo{pages}{195107} (\bibinfo{year}{2010}),
  \urlprefix\url{http://link.aps.org/doi/10.1103/PhysRevB.81.195107}.

\bibitem[{\citenamefont{Haule}(2018)}]{JPSJ_Haule}
\bibinfo{author}{\bibfnamefont{K.}~\bibnamefont{Haule}},
  \bibinfo{journal}{Journal of the Physical Society of Japan}
  \textbf{\bibinfo{volume}{87}}, \bibinfo{pages}{041005}
  (\bibinfo{year}{2018}), \eprint{https://doi.org/10.7566/JPSJ.87.041005},
  \urlprefix\url{https://doi.org/10.7566/JPSJ.87.041005}.

\bibitem[{web()}]{webpage}
\emph{\bibinfo{title}{{DFT + Embedded DMFT} functional}},
  \bibinfo{howpublished}{\url{http://hauleweb.rutgers.edu/tutorials/}},
  \eprint{, where for the DFT part we are using the WIEN2k code: P. Blaha, K.
  Schwarz, G. K. H. Madsen, D. Kvasnicka, J. Luitz, R. Laskowski, F. Tran and
  L. D. Marks, “WIEN2k, An Augmented Plane Wave + Local Orbitals Program for
  Calculating Crystal Properties” (Karlheinz Schwarz, Techn. Universit{\"a}t
  Wien, Austria), 2018. ISBN 3-9501031-1-2}.

\bibitem[{\citenamefont{Kittel}(2005)}]{Kittel8}
\bibinfo{author}{\bibfnamefont{C.}~\bibnamefont{Kittel}},
  \emph{\bibinfo{title}{Introduction to Solid State Physics}}
  (\bibinfo{publisher}{Wiley}, \bibinfo{year}{2005}), \bibinfo{edition}{eighth}
  ed.

\bibitem[{\citenamefont{Ziman}(1972)}]{Ziman72}
\bibinfo{author}{\bibfnamefont{J.~M.} \bibnamefont{Ziman}},
  \emph{\bibinfo{title}{Principles of Theory of Solids}}
  (\bibinfo{publisher}{Cambridge University Press}, \bibinfo{year}{1972}),
  \bibinfo{edition}{3rd} ed.

\bibitem[{\citenamefont{Ketterson}(2016)}]{Ketterson16}
\bibinfo{author}{\bibfnamefont{J.~B.} \bibnamefont{Ketterson}},
  \emph{\bibinfo{title}{The Physics of Solids}} (\bibinfo{publisher}{Oxford
  University Press}, \bibinfo{year}{2016}), \bibinfo{edition}{1st} ed.

\bibitem[{\citenamefont{White et~al.}(2013)\citenamefont{White, Janoschek,
  Kanchanavatee, Huang, Shu, Jang, D.Y.~Tutun, Lum, Baumbach, and
  Maple}}]{thermoelectric_capabilities1}
\bibinfo{author}{\bibfnamefont{B.}~\bibnamefont{White}},
  \bibinfo{author}{\bibfnamefont{M.}~\bibnamefont{Janoschek}},
  \bibinfo{author}{\bibfnamefont{N.}~\bibnamefont{Kanchanavatee}},
  \bibinfo{author}{\bibfnamefont{K.}~\bibnamefont{Huang}},
  \bibinfo{author}{\bibfnamefont{L.}~\bibnamefont{Shu}},
  \bibinfo{author}{\bibfnamefont{S.}~\bibnamefont{Jang}},
  \bibinfo{author}{\bibfnamefont{J.~H.} \bibnamefont{D.Y.~Tutun}},
  \bibinfo{author}{\bibfnamefont{I.}~\bibnamefont{Lum}},
  \bibinfo{author}{\bibfnamefont{R.}~\bibnamefont{Baumbach}}, \bibnamefont{and}
  \bibinfo{author}{\bibfnamefont{M.}~\bibnamefont{Maple}},
  \emph{\bibinfo{title}{Thermoelectric properties of correlated electron
  systems Ln$_3$Pt$_4$Ge$_6$ and LnPt$_4$Ge$_{12}$(Ln = Ce, Pr) and
  non-centrosymmetric X$_2$T$_{12}$P$_7$(X = Yb, Hf and T = Fe, Co). In: Zlatic
  V., Hewson A. (eds) New Materials for Thermoelectric Applications: Theory and
  Experiment. NATO Science for Peace and Security Series B: Physics and
  Biophysics. Springer, Dordrecht}} (\bibinfo{year}{2013}).

\bibitem[{\citenamefont{Rourke and Julian}(2012)}]{Haas_van_ROURKE2012324}
\bibinfo{author}{\bibfnamefont{P.}~\bibnamefont{Rourke}} \bibnamefont{and}
  \bibinfo{author}{\bibfnamefont{S.}~\bibnamefont{Julian}},
  \bibinfo{journal}{Computer Physics Communications}
  \textbf{\bibinfo{volume}{183}}, \bibinfo{pages}{324 } (\bibinfo{year}{2012}),
  ISSN \bibinfo{issn}{0010-4655},
  \urlprefix\url{http://www.sciencedirect.com/science/article/pii/S001046551100347X}.

\bibitem[{\citenamefont{Blake et~al.}(2015)\citenamefont{Blake, Watson,
  McCollam, Kasahara, Johnson, Narayanan, Pascut, Haule, Kiryukhin, Yamashita
  et~al.}}]{Haas_van_PhysRevB.91.121105}
\bibinfo{author}{\bibfnamefont{S.~F.} \bibnamefont{Blake}},
  \bibinfo{author}{\bibfnamefont{M.~D.} \bibnamefont{Watson}},
  \bibinfo{author}{\bibfnamefont{A.}~\bibnamefont{McCollam}},
  \bibinfo{author}{\bibfnamefont{S.}~\bibnamefont{Kasahara}},
  \bibinfo{author}{\bibfnamefont{R.~D.} \bibnamefont{Johnson}},
  \bibinfo{author}{\bibfnamefont{A.}~\bibnamefont{Narayanan}},
  \bibinfo{author}{\bibfnamefont{G.~L.} \bibnamefont{Pascut}},
  \bibinfo{author}{\bibfnamefont{K.}~\bibnamefont{Haule}},
  \bibinfo{author}{\bibfnamefont{V.}~\bibnamefont{Kiryukhin}},
  \bibinfo{author}{\bibfnamefont{T.}~\bibnamefont{Yamashita}},
  \bibnamefont{et~al.}, \bibinfo{journal}{Phys. Rev. B}
  \textbf{\bibinfo{volume}{91}}, \bibinfo{pages}{121105}
  (\bibinfo{year}{2015}),
  \urlprefix\url{https://link.aps.org/doi/10.1103/PhysRevB.91.121105}.

\bibitem[{\citenamefont{Brasse et~al.}(2013)\citenamefont{Brasse, Chioncel,
  Kune\ifmmode~\check{s}\else \v{s}\fi{}, Bauer, Regnat, Blum, Wurmehl,
  Pfleiderer, Wilde, and Grundler}}]{Haas_van_PhysRevB.88.155138}
\bibinfo{author}{\bibfnamefont{M.}~\bibnamefont{Brasse}},
  \bibinfo{author}{\bibfnamefont{L.}~\bibnamefont{Chioncel}},
  \bibinfo{author}{\bibfnamefont{J.}~\bibnamefont{Kune\ifmmode~\check{s}\else
  \v{s}\fi{}}}, \bibinfo{author}{\bibfnamefont{A.}~\bibnamefont{Bauer}},
  \bibinfo{author}{\bibfnamefont{A.}~\bibnamefont{Regnat}},
  \bibinfo{author}{\bibfnamefont{C.~G.~F.} \bibnamefont{Blum}},
  \bibinfo{author}{\bibfnamefont{S.}~\bibnamefont{Wurmehl}},
  \bibinfo{author}{\bibfnamefont{C.}~\bibnamefont{Pfleiderer}},
  \bibinfo{author}{\bibfnamefont{M.~A.} \bibnamefont{Wilde}}, \bibnamefont{and}
  \bibinfo{author}{\bibfnamefont{D.}~\bibnamefont{Grundler}},
  \bibinfo{journal}{Phys. Rev. B} \textbf{\bibinfo{volume}{88}},
  \bibinfo{pages}{155138} (\bibinfo{year}{2013}),
  \urlprefix\url{https://link.aps.org/doi/10.1103/PhysRevB.88.155138}.

\end{thebibliography}

\end{document}